# Design of 5G Full Dimension Massive MIMO Systems

Qurrat-Ul-Ain Nadeem, *Student Member, IEEE*, Abla Kammoun, *Member, IEEE*, Mérouane Debbah, *Fellow, IEEE*, and Mohamed-Slim Alouini, *Fellow, IEEE*

*Abstract*—This paper discusses full-dimension multiple-input-multiple-output (FD-MIMO) technology, which is currently an active area of research and standardization in wireless communications for evolution toward Fifth Generation (5G) cellular systems. FD-MIMO utilizes an active antenna system (AAS) with a 2-D planar array structure that not only allows a large number of antenna elements to be packed within feasible base station form factors, but also provides the ability of adaptive electronic beamforming in the 3-D space. However, the compact structure of large-scale planar arrays drastically increases the spatial correlation in FD-MIMO systems. In order to account for its effects, the generalized spatial correlation functions for channels constituted by individual elements and overall antenna ports in the AAS are derived. Exploiting the quasi-static channel covariance matrices of users, the problem of determining the optimal downtilt weight vector for antenna ports, which maximizes the minimum signal-to-interference ratio of a multi-user multiple-input-single-output system, is formulated as a fractional optimization problem. A quasi-optimal solution is obtained through the application of semi-definite relaxation and Dinkelbach's method. Finally, the user-specific elevation beamforming scenario is devised, which offers significant performance gains as confirmed through simulations. These results have direct application in the analysis of 5G FD-MIMO systems.

## I. INTRODUCTION

MULTIPLE-INPUT multiple-output (MIMO) technology has remained a subject of interest in the last two decades due to its ability to cope with the increase in the

Manuscript received April 22, 2017; revised August 29, 2017; accepted October 6, 2017. Date of publication October 13, 2017; date of current version February 14, 2018. The work of Q.-U.-A. Nadeem, A. Kammoun and M.-S. Alouini was supported by a CRG4 grant from the Office of Competitive Research Funding (OCRF) at KAUST. The work of Mérouane Debbah was supported by the ERC Starting Grant 305123 MORE (Advanced Mathematical Tools for Complex Network Engineering). This paper was presented at the IEEE Global Communications Conference, Washington, DC, USA, in December 2016. The associate editor coordinating the review of this paper and approving it for publication was X. Yuan. *(Corresponding author: Qurrat-Ul-Ain Nadeem.)*

Q.-U.-A. Nadeem, A. Kammoun, and M.-S. Alouini are with the Computer, Electrical, and Mathematical Sciences and Engineering Division, King Abdullah University of Science and Technology, Thuwal 23955-6900, Saudi Arabia (e-mail: qurratulain.nadeem@kaust.edu.sa; abla.kammoun@kaust.edu.sa; slim.alouini@kaust.edu.sa).

M. Debbah is with Centrale Supélec, 91192 Gif-sur-Yvette, France, and also with the Mathematical and Algorithmic Sciences Laboratory, Huawei France R&D, 92100 Paris, France (e-mail: merouane.debbah@huawei.com; merouane.debbah@centralesupelec.fr).

Color versions of one or more of the figures in this paper are available online at http://ieeexplore.ieee.org.

Digital Object Identifier 10.1109/TCOMM.2017.2762685

wireless data traffic and improve the reliability of wireless systems. In order to be compatible with the existing 3rd Generation Partnership Project (3GPP) Long Term Evolution (LTE) standard, most of the existing MIMO implementations consider the deployment of fewer than ten linearly placed antennas at the base station (BS) [1]. The corresponding improvement in spectral efficiency, although important, is still relatively modest and can be vastly improved by scaling up these systems by possibly orders of magnitude. This has led to the introduction of massive MIMO systems, that serve many users in the same time-frequency resource using low-complexity signal processing methods [2]–[4].

While massive MIMO technology is a key enabler for next generation cellular systems, there are still many practical challenges down the road to its successful deployment [2], [4]. One main challenge is that the number of antennas that can be equipped at the top of BS towers is limited by the BS form factors and operating LTE carrier frequencies. To circumvent this difficulty, the 3GPP has proposed the use of a 2D uniform planar array which can be readily installed in practice as compared to the conventional uniform linear array (ULA) [5]–[7]. For example, a $16 \times 16$ uniform planar array with half wavelength inter-element spacing occupies about 1m×1m space at the typical LTE carrier frequency of 2.5 GHz. By contrast, about 15m spacing is required in the horizontal direction to install a linear array of 256 antennas, which is not feasible given the limited installation space at BS towers.

In addition to the emergence of large-scale antenna arrays, the cell site architecture itself has evolved in the last decade from the one wherein the base transceiver station (BTS) equipment is located away from the passive antenna element array, to the one wherein the analogue portion of the BTS, comprising of amplifiers and phase shifters, is located in the remote radio head closer to the passive antenna elements. The next stage is the integration of the active transceiver unit into the passive antenna element array, resulting in an active antenna system (AAS) [8], [9]. The AAS can support adaptive electronic beamforming by controlling the phase and amplitude weights applied to individual antenna elements. The use of an AAS with a 2D planar array structure results in full-dimension (FD) MIMO, which was identified as a promising technology for 5G cellular systems in the 3GPP Release-12 workshop in 2012 [10]. Follow-up study items are under completion [11] and formal standardization is being done in Release-13.

FD-MIMO has three important distinguishing features as compared to the conventional LTE systems. Firstly, the number of antennas supported within feasible BS form factors, indicated in [12, Table II], has increased due to the 2D planar structure of the antenna arrays. Secondly, the 2D arrangement of active antenna elements provides the ability of adaptive electronic beam control over both the elevation and the traditional azimuth dimensions [13], [14]. The radio resource is organized on the basis of antenna ports, where each port is mapped to a group of physical antenna elements arranged in the vertical direction [7]. Controlling the phase and amplitude weights applied to these elements allows for the dynamic adaptation of the vertical dimension of the antenna port radiation pattern. This technology, referred to as elevation beamforming, enables more directed and spatially separated transmissions to a large number of users [8], [9], [15]–[17]. Thirdly, the FD-MIMO architecture lends itself to a hybrid digital analog implementation, where the multi-user (MU) MIMO precoding stage is implemented in the baseband and the elevation beamforming stage is implemented in the analog domain [18], [19].

The hybrid implementation helps reduce the channel state information (CSI) feedback overhead, which is prohibitively high when implementing fully digital MU-MIMO beamforming techniques in massive MIMO systems. In an attempt to reduce this feedback overhead, the authors in [20] exploit the sparsity in millimeter wave (mmWave) channels as well as the notion of compressive sensing to design a low-complexity beam selection method that does not require explicit channel estimation. Other works [18], [21] employ channel dimensionality reduction techniques before the MU precoding stage. The FD-MIMO active antenna arrays realize this channel dimension reduction in the elevation beamforming stage, where the downtilt weight vectors are used to group antenna elements into a reduced number of antenna ports. The digital precoding stage then requires the estimation of the reduced-dimension effective channel.

The additional control over the elevation dimension in FD-MIMO systems enables a variety of strategies such as sector-specific and user-specific elevation beamforming, and cell splitting [13], [22], [23]. The authors in [17] used lab and field trials to show that 3D beamforming can achieve significant performance gains in real indoor and outdoor deployments, by adapting the vertical dimension of the antenna port radiation pattern at the BS individually for each user according to its location. Some 3D beamforming designs were proposed in [15] for a single user multiple-input-single-output (MISO) system, wherein the authors used the approximate antenna port radiation pattern expressions from [24], [25] to find the optimal downtilt angle. This approach, also used in works dealing with multi-user scenarios [23], [26], [27], abstracts the role played by physical elements constituting an antenna port in performing the downtilt by approximating the vertical radiation pattern of each port by a narrow beam in the elevation. The actual radiation pattern depends on the number of elements constituting the port, their patterns, relative positions and corresponding weights [7], [11]. More sophisticated elevation beamforming methods need to be developed that depend directly on the underlying structure of the port.

This paper discusses the design of the 2D active antenna array for FD-MIMO implementation. The array constitutes of columns of active antenna elements, where each column is referred to as an antenna port and is fed with a corresponding downtilt weight vector to steer the vertical radiation pattern of that port in the targeted direction. In our study of FD-MIMO systems, we make three main contributions which are summarized as follows.

- Compact structure of large-scale antenna arrays drastically increases the spatial correlation in FD-MIMO systems, making it imperative to account for it in the design of elevation beamforming algorithms. The existing studies on antenna correlation consider passive antenna elements arranged in the azimuth plane only [28]–[32]. In this work, we derive the exact spatial correlation function (SCF) for the 3D channels constituted by individual antenna elements in the 2D AAS by expanding the array response of each antenna element using the spherical harmonic expansion (SHE) of plane waves. The SCF is presented in **Theorem 1**. The analysis follows the guidelines in [33], where we derived the SCF for a ULA of antenna ports using the approximate antenna port radiation pattern expression from [24]. The final analytical expression depends on the angular parameters and the geometry of the array through the Fourier series (FS) coefficients of the power spectra. The correlation between the antenna ports is then expressed as a function of the correlation matrix of the elements constituting the ports and their corresponding downtilt weight vectors.

- The main objective of this article is to study the performance benefits of FD-MIMO techniques by devising efficient elevation beamforming algorithms that optimize the downtilt weight vectors of the antenna ports, utilizing the quasi-static channel correlation matrices of the users obtained using the derived SCF. The weight vectors that maximize the signal-to-noise ratio (SNR) of a single user MISO system are presented in **Theorem 2**. The downlink of a multi-user MISO system is studied next under the assumption that all antenna ports transmit using the same downtilt weight vector. The problem of determining this vector that maximizes the minimum signal-to-interference ratio (SIR) of the system is formulated as a fractional optimization problem, for which a quasi-optimal solution is obtained through the application of semi-definite relaxation and Dinkelbach's method, as summarized in Algorithm 1. This scenario is referred to as single downtilt beamforming (SDB).

- Finally, we devise a user-group specific single downtilt beamforming (UG-SDB) scenario in Algorithm 2, wherein the user population is partitioned into groups based on the users' quasi-static channel correlation matrices and each user-group is served by a subset of antenna ports that transmit using a quasi-optimal downtilt weight vector. Nam *et al.* have proposed two user-grouping schemes based on the users' channel correlation matrices in [21] - the K-means clustering and the fixed quantization method. Our user-grouping method is motivated by the latter. Simulations results show that even the SDB scenario yields significant performance gains when compared to the existing elevation beamforming methods in literature, while user-group specific beamforming further improves the system performance.

The rest of this article is organized as follows. Section II discusses the design of the 2D AAS for FD-MIMO implementation and presents the corresponding 3D channel model. In section III, the exact SCFs for channels constituted by the individual antenna elements and the overall antenna ports are derived. Section IV presents the elevation beamforming algorithms that optimize the downtilt weight vectors of the antenna ports in the single and multi-user MISO settings. User-group specific single downtilt beamforming scenario is devised in section V. Section VI provides simulation results and finally, in section VII some concluding remarks are drawn.

## II. ANTENNA CONFIGURATION AND 3D CHANNEL MODEL

The idea of exploiting the elevation domain of the channel for performance optimization has led to the development of FD-MIMO systems. The recent reports on FD-MIMO envision that an AAS, utilizing a large number of antenna elements arranged in a 2D planar array structure, can be designed to realize spatially separated transmission links to a large number of users. In this section, we introduce this 2D AAS and outline the corresponding FD-MIMO channel model.

### A. Active Antenna Array for FD-MIMO

In order to realize the performance benefits of FD-MIMO techniques, an efficient implementation of an AAS with a 2D planar array structure is a key requirement. The AAS is an advanced BS technology, which integrates the active transceiver unit array into the passive antenna element array, allowing the gain, beamwidth and downtilt of the transmit beam to be controlled adaptively by active electronic components connected directly to each element [8], [9]. These active antenna elements should be placed in both the vertical and horizontal directions to provide the ability of adaptive electronic beamforming in the elevation and the traditional azimuth dimensions.

The 3GPP proposes the organization of the radio resource on the basis of antenna ports, where each port is mapped to a group of physical antenna elements arranged in the vertical domain. The elements in a port carry the same signal and are fed with corresponding downtilt weights to focus the wavefront in the direction of the targeted user. The structure of a typical antenna port comprising of $N_E$ antenna elements is shown in Fig. 1. There are $N_{BS}$ such ports placed at equidistant positions in the $\hat{\mathbf{e}}_y$ direction, where the downtilt angle, $\theta_{tilt}$, of the radiation pattern of every port is controlled through the applied weights $w_k(\theta_{tilt})$, $k = 1, \ldots, N_E$. The resulting configuration for vertically polarized antenna elements is shown in Fig. 2. This generic AAS architecture therefore takes a 2D planar array structure for antenna elements, which is more feasible in terms of form factor as compared to the ULA configuration.

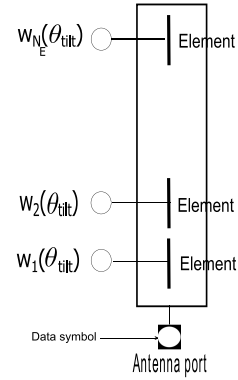

Fig. 1. Antenna port.

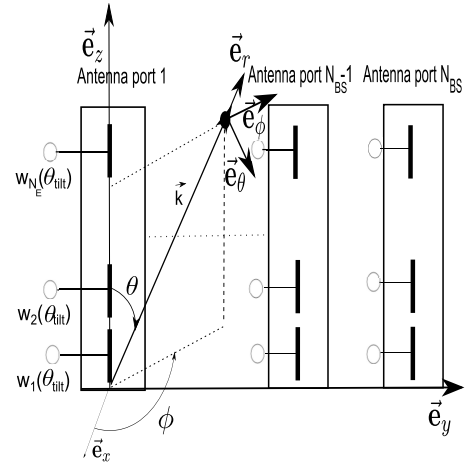

Fig. 2. Active antenna array.

### B. Antenna Element Approach towards 3D Channel Modeling

Preliminary studies on 3D channel modeling [14], [33] consider channels between the overall antenna ports rather than between the physical elements constituting these ports and use the approximate antenna port radiation pattern expression from [24], [25]. In theory, the radiation pattern of an antenna port depends on the number of elements within it, their positions, individual patterns and corresponding weights. In other words, it is a superposition of the element radiation pattern and the array factor for that port, where the element radiation pattern is given by [6], [7],

$$A_E(\phi, \theta) = G_{E,max} - \min\{-(A_{E,H}(\phi) + A_{E,V}(\theta)), A_m\},$$
(1)

where,

$$A_{E,H}(\phi) = -\min\left[12\left(\frac{\phi}{\phi_{3dB}}\right)^2, A_m\right] \text{dB},$$
(2)

$$A_{E,V}(\theta) = -\min\left[12\left(\frac{\theta - 90^o}{\theta_{3dB}}\right)^2, SLA_v\right] \text{dB},$$
(3)

where $\phi$ and $\theta$ denote the azimuth and elevation angles respectively, $A_{E,H}(\phi)$ and $A_{E,V}(\theta)$ are the radiation patterns in the horizontal and vertical directions respectively, $G_{E,max}$ is the maximum directional element gain, $\phi_{3dB}$ and $\theta_{3dB}$ are the half power beamwidths in the azimuth and elevation domains respectively, $A_m$ is the maximum attenuation and $SLA_v$ is the vertical side lobe attenuation level. The global field pattern of a vertically polarized antenna element in linear scale is $\sqrt{A_E(\phi, \theta)|_{\text{lin}}} = \sqrt{10^{\frac{G_{E,max}}{10}}} g_E(\phi, \theta)$, where $g_E(\phi, \theta) \approx g_{E,H}(\phi) g_{E,V}(\theta)$ with,

$$g_{E,H}(\phi) = \exp\left(-1.2 \left(\frac{\phi}{\phi_{3dB}}\right)^2 \ln 10\right), \tag{4}$$

$$g_{E,V}(\theta) = \exp\left(-1.2 \left(\frac{\theta - 90^o}{\theta_{3dB}}\right)^2 \ln 10\right). \tag{5}$$

The overall array radiation pattern is a function of this individual element radiation pattern and the array factor matrix, $\mathbf{A}$, for the AAS given by [7],

$$\mathbf{A} = \mathbf{W} \circ \mathbf{V}, \tag{6}$$

where $\circ$ is the Hadamard product, $\mathbf{V}$ is a $N_E \times N_{BS}$ matrix containing the array responses of the individual radiation elements with each entry given as,

$$[\mathbf{V}]_{z,s} = \exp(i\mathbf{k}.\mathbf{x}_{z,s}), \quad z = 1, \ldots, N_E, \ s = 1, \ldots, N_{BS}, \tag{7}$$

where $.$ is the scalar dot product, $\mathbf{x}_{z,s}$ is the location vector of the $z^{th}$ antenna element in the $s^{th}$ Tx antenna port, and $\mathbf{k}$ is the Tx wave vector, where $\mathbf{k} = \frac{2\pi}{\lambda} \hat{\mathbf{v}}$, with $\hat{\mathbf{v}}$ being the unit wave vector. For the configuration shown in Fig. 2, every entry of $\mathbf{V}$ will have a form given by,

$$[\mathbf{V}]_{z,s}(\phi, \theta)$$
$$= \exp\left(i 2\pi \left((s-1)\frac{d_y}{\lambda}\sin\phi\sin\theta + (z-1)\frac{d_z}{\lambda}\cos\theta\right)\right), \tag{8}$$

where $d_y$ is the horizontal separation between the antenna ports and $d_z$ is the vertical separation between the antenna elements, with the phase reference at the origin.

Also $\mathbf{W}$ is a $N_E \times N_{BS}$ matrix comprising of the weights to be applied to the individual radiation elements. The 3GPP TR36.873 [6] and TR37.840 [7] assume the magnitude of the weights to be identical for each radiation element, while the phase is used to implement the electrical downtilt. Denoting the $(z, s)^{th}$ entry of $\mathbf{W}$ as $w_s^z$, the 3GPP proposed expression for $w_s^z$ is,

$$w_s^z = \frac{1}{\sqrt{N_E}} \exp\left(-i 2\pi (z-1)\frac{d_z}{\lambda}\cos\theta_{tilt_s}\right), \tag{9}$$

where $\theta_{tilt_s}$ is the downtilt angle for the $s^{th}$ port, defined between $0^o$ and $180^o$. Denoting the $(z, s)^{th}$ entry of $\mathbf{V}$ as $v_s^z(\phi, \theta)$, the small-scale 3D channel constituted by the BS antenna port $s$ is given as,

$$[\mathbf{h}]_s = \sum_{(z \in \text{port } s)=1}^{N_E} w_s^z \sum_{n=1}^{N} \alpha_n \sqrt{g_E(\phi_n, \theta_n)} v_s^z(\phi_n, \theta_n), \tag{10}$$

$$= \mathbf{w}_s^T \sum_{n=1}^{N} \alpha_n \sqrt{g_E(\phi_n, \theta_n)} \mathbf{v}_s(\phi_n, \theta_n), \quad s = 1, \ldots, N_{BS}, \tag{11}$$

where $N$ is the number of propagation paths, $\phi_n$ and $\theta_n$ are the azimuth and elevation angle of departure (AoD) of the $n^{th}$ path respectively, and $\alpha_n \sim$ i.i.d $\mathcal{CN}(0, \frac{1}{N})$ is the amplitude of the $n^{th}$ path. Also $\mathbf{w}_s$ is the weight vector for the $s^{th}$ antenna port, given by the $s^{th}$ column of $\mathbf{W}$, and $\mathbf{v}_s(\phi_n, \theta_n)$ is the $s^{th}$ column of $\mathbf{V}$. The channel with a given antenna port is therefore a weighted sum of the channels with the $N_E$ elements inside it.

## III. Spatial Correlation Function

The compact structure of FD-MIMO arrays dramatically increases the spatial correlation between the antenna ports, which is a function of the correlation between the elements constituting these ports. In order to realistically evaluate the performance of FD-MIMO techniques, we derive generalized analytical expressions for these SCFs, considering arbitrary AoD distributions and antenna patterns.

### A. Spatial Correlation Function for Antenna Elements

Using the antenna element radiation pattern expressions given in (4) and (5) and the array response expression given in (8), and for $\alpha_n \sim$ i.i.d $\mathcal{CN}(0, \frac{1}{N})$ random variables, the SCF for the channels constituted by $(z, s)^{th}$ and $(z', s')^{th}$ antenna elements in the AAS, where $z, z' = 1, \ldots, N_E$, $s, s' = 1, \ldots, N_{BS}$, is expressed as,

$$\rho_E(s - s', z - z')$$
$$= \mathbb{E}[g_E(\phi, \theta) v_s^z(\phi, \theta) v_{s'}^{z'*}(\phi, \theta)],$$
$$= \mathbb{E}\Big[g_E(\phi, \theta) \exp\Big(i 2\pi \Big[\frac{d_y}{\lambda}(s - s')\sin\phi\sin\theta$$
$$+ \frac{d_z}{\lambda}(z - z')\cos\theta\Big]\Big)\Big]. \tag{12}$$

In order to obtain an analytical expression for $\rho_E(s - s', z - z')$, an approach similar to the one used in [33] will be adopted. The main difference is that here we work with a 2D antenna array instead of a linear array, which makes the analysis more involved. As a starting point, let $Z_y = (s - s')\frac{d_y}{\lambda}$ and $Z_z = (z - z')\frac{d_z}{\lambda}$, and define $Z = \sqrt{Z_y^2 + Z_z^2}$ and,

$$\beta = \begin{cases} 0, & \text{if } s - s' = 0 \& z - z' = 0, \\ \arctan\left(\frac{Z_y}{Z_z}\right), & \text{if } s - s' > 0 \& z - z' \geq 0, \\ \pi + \arctan\left(\frac{Z_y}{Z_z}\right), & \text{if } s - s' \geq 0 \& z - z' < 0. \end{cases} \tag{13}$$

With these definitions and reformulations, (12) can be expressed as,

$$\rho_E(s - s', z - z') = \mathbb{E}\Big[g_E(\phi, \theta) \exp\Big(i 2\pi Z \Big[\cos\theta \cos\beta$$
$$+ \sin\theta \sin\beta \cos\Big(\phi - \frac{\pi}{2}\Big)\Big]\Big)\Big]. \tag{14}$$

Observe that $\cos\theta\cos\beta + \sin\theta\sin\beta\cos\left(\phi - \frac{\pi}{2}\right)$ is the dot product of $\hat{\mathbf{v}}$ with spherical coordinates $(\phi, \theta)$ and $\hat{\mathbf{x}}$ with spherical coordinates $(\pi/2, \beta)$, where $\mathbf{x}$ is the location vector between the $(z, s)^{th}$ and $(z', s')^{th}$ antenna elements in the AAS. Exploiting results on the SHE of plane waves and the Legendre addition theorem provided in [33] will expand (14) as,

$$
\begin{aligned}
\rho_E(s-s',&\, z-z') \\
&= \mathbb{E}\Big[ g_E(\phi,\theta) \sum_{n=0}^{\infty} i^n (2n+1) j_n (2\pi Z) \\
&\quad \times \Big( P_n(\cos\beta) P_n(\cos\beta) + 2 \sum_{m=1}^{n} \frac{(n-m)!}{(n+m)!} P_n^m(\cos\theta) \\
&\quad \times P_n^m(\cos\beta) \cos\Big(m\Big(\phi - \frac{\pi}{2}\Big)\Big)\Big)\Big].
\end{aligned}
\tag{15}
$$

Next we systematically expand $\rho_E(s-s', z-z')$ by defining $\bar{P}_n^m(\mathrm{x}) = \sqrt{(n+\frac{1}{2})\frac{(n-m)!}{(n+m)!}} P_n^m(\mathrm{x})$ and using the decomposition $g_E(\phi,\theta) \approx g_{E,H}(\phi) g_{E,V}(\theta)$ to yield (16), as shown at the top of the next page.

The random variables, AoDs, appear as arguments of Legendre polynomials in (16) and it is important to have a general representation for these polynomials to facilitate the analytical development of the expectation terms. For this purpose, we use the trigonometric expansion of Legendre polynomials summarized in the following Lemma [34].

**Lemma 1.** For non-negative integers $n$ and $m$,

$$
P_{2n}(\cos x) = p_n^2 + 2\sum_{k=1}^{n} p_{n-k} p_{n+k} \cos(2kx),
$$

$$
P_{2n-1}(\cos x) = 2\sum_{k=1}^{n} p_{n-k} p_{n+k-1} \cos((2k-1)x),
$$

$$
\bar{P}_{2n}^{2m}(\cos x) = \sum_{k=0}^{n} c_{2n,2k}^{2m} \cos(2kx),
$$

$$
\bar{P}_{2n}^{2m-1}(\cos x) = \sum_{k=1}^{n} d_{2n,2k}^{2m-1} \sin(2kx),
$$

$$
\bar{P}_{2n-1}^{2m}(\cos x) = \sum_{k=1}^{n} c_{2n-1,2k-1}^{2m} \cos((2k-1)x),
$$

$$
\bar{P}_{2n-1}^{2m-1}(\cos x) = \sum_{k=1}^{n} d_{2n-1,2k-1}^{2m-1} \sin((2k-1)x),
\tag{17}
$$

where $p_n$, $c_{2n,2k}^{2m}$, $c_{2n-1,2k-1}^{2m}$, $d_{2n,2k}^{2m-1}$ and $d_{2n-1,2k-1}^{2m-1}$ are given by the recursion relations in [34].

These expansions assist in the expression of $\rho_E(s-s', z-z')$ in terms of the FS coefficients of the power azimuth spectrum (PAS) and the power elevation spectrum (PES) defined as [35],

$$
\mathrm{PAS}_E(\phi) = g_{E,H}(\phi) p_\phi(\phi),
\tag{18}
$$

$$
\mathrm{PES}_E(\theta) = g_{E,V}(\theta) p_\theta(\theta),
\tag{19}
$$

where the angular power density functions, $p_\phi(\phi)$ and $p_\theta(\theta)$ equal $f_\phi(\phi)$ and $\frac{f_\theta(\theta)}{\sin(\theta)}$ respectively, with $f_\phi(\phi)$ and $f_\theta(\theta)$ being the probability density functions of the azimuth and elevation angles. The FS coefficients, $a_\phi(m)$, $b_\phi(m)$, $a_\theta(k)$ and $b_\theta(k)$, for the power spectra are defined as,

$$
a_\phi(m) = \frac{1}{\pi} \int_{-\pi}^{\pi} \mathrm{PAS}_E(\phi) \cos(m\phi) d\phi,
\tag{20}
$$

$$
b_\phi(m) = \frac{1}{\pi} \int_{-\pi}^{\pi} \mathrm{PAS}_E(\phi) \sin(m\phi) d\phi,
\tag{21}
$$

$$
a_\theta(k) = \frac{1}{\pi} \int_{0}^{\pi} \mathrm{PES}_E(\theta) \cos(k\theta) d\theta,
\tag{22}
$$

$$
b_\theta(k) = \frac{1}{\pi} \int_{0}^{\pi} \mathrm{PES}_E(\theta) \sin(k\theta) d\theta.
\tag{23}
$$

The final analytical expression for the SCF is presented in **Theorem 1**.

**Theorem 1.** For an AAS with a 2D uniform planar array of antenna elements with arbitrary antenna patterns and for arbitrary angular distributions, such that $\phi \in [-\pi, \pi]$ and $\theta \in [0, \pi]$, the 3D SCF for the channels constituted by $(z, s)^{th}$ and $(z', s')^{th}$ antenna elements is given by (24), as shown at the top of the next page, for $(s - s' = 0 \ \& \ z - z' = 0)$, $(s - s' > 0 \ \& \ z - z' \geq 0)$ and $(s - s' \geq 0 \ \& \ z - z' < 0)$. For other cases, $\rho_E(s - s', z - z')$ is computed as $\rho_E(s - s', z - z') = \rho_E(s' - s, z' - z)^*$ for $(s - s' < 0 \ \& \ z - z' \leq 0)$, and $\rho_E(s - s', z - z') = \rho_E(s' - s, z' - z)^*$ for $(s - s' \leq 0 \ \& \ z - z' > 0)$.

Using the approximation provided in [33], the summations over $n$ can be truncated to $N_0 = 10$ terms, such that the truncation error in the correlation between adjacent antenna elements in the AAS with $.5\lambda$ spacing is bounded by $\sim 0.5\%$.

**Remark 1.** The correlation analysis can be extended to frequency-selective channels using orthogonal frequency-division multiplexing (OFDM) signaling. Specifically, a $V$ sub-carriers MISO-OFDM system with proper cyclic prefix extension converts the frequency selective counter-part of (10) into $V$ flat-fading channels, where the channel transfer function corresponding to the $v^{th}$ sub-carrier and the $z^{th}$ element in the $s^{th}$ port is given as [36],

$$
H_{v,z,s} = \sum_{n=1}^{N} h_{z,s,n} \exp\left(-i 2\pi \frac{v}{V} n\right),
\tag{25}
$$

where,

$$
h_{z,s,n} = \alpha_n \sqrt{g_E(\phi_n, \theta_n)} v_s^z(\phi_n, \theta_n).
\tag{26}
$$

The SCF for the flat-fading channels constituted by $(z, s)^{th}$ and $(z', s')^{th}$ antenna elements in the AAS corresponding to the $v^{th}$ sub-carrier is expressed as,

$$
\begin{aligned}
\rho_{E,v}(s-s', z-z') = \sum_{n=1}^{N} \sum_{n'=1}^{N} \mathbb{E}[&\alpha_n \alpha_{n'}^* \sqrt{g_E(\phi_n, \theta_n)} \\
&\times \sqrt{g_E(\phi_{n'}, \theta_{n'})} v_s^z(\phi_n, \theta_n) v_{s'}^{z'*}(\phi_{n'}, \theta_{n'}) \\
&\times \exp\left(-i 2\pi \left(\frac{v}{V}\right)(n-n')\right)].
\end{aligned}
\tag{27}
$$

For $\alpha_n \sim$ i.i.d $\mathcal{CN}(0, \frac{1}{N})$ and $(\phi_n, \theta_n)$, $n = 1, \ldots, N$, also being i.i.d, the SCF in (27) can be expressed as (12). The final analytical expression will therefore be identical for the $V$ sub-carriers and given by (24).

### B. Spatial Correlation Function for Antenna Ports

In FD-MIMO techniques, the radio resource is organized on the basis of antenna ports, where each port is used to transmit data symbols at a particular value of the downtilt angle. It is therefore important to characterize the correlation between

$$\rho_E(s-s', z-z') = \mathbb{E}[g_E(\phi, \theta)] j_0(2\pi Z) + \sum_{n=1}^{\infty} (-1)^n (4n+1) j_{2n}(2\pi Z) P_{2n}(\cos\beta) \mathbb{E}[P_{2n}(\cos\theta) g_{E,V}(\theta)]$$

$$\times \mathbb{E}[g_{E,H}(\phi)] - \sum_{n=1}^{\infty} i(-1)^n (4n-1) j_{2n-1}(2\pi Z) P_{2n-1}(\cos\beta) \mathbb{E}[P_{2n-1}(\cos\theta) g_{E,V}(\theta)] \mathbb{E}[g_{E,H}(\phi)]$$

$$+ \sum_{n=1}^{\infty} 4(-1)^n j_{2n}(2\pi Z) \left( \left( \sum_{m=1}^{n} (-1)^m \tilde{P}_{2n}^{2m}(\cos\beta) \mathbb{E}[\tilde{P}_{2n}^{2m}(\cos\theta) g_{E,V}(\theta)] \mathbb{E}[\cos(2m\phi) g_{E,H}(\phi)] \right) \right.$$

$$\left. - \left( \sum_{m=1}^{n} (-1)^m \tilde{P}_{2n}^{2m-1}(\cos\beta) \mathbb{E}[\tilde{P}_{2n}^{2m-1}(\cos\theta) g_{E,V}(\theta)] \mathbb{E}[\sin((2m-1)\phi) g_{E,H}(\phi)] \right) \right) + \sum_{n=1}^{\infty} 4i(-1)^n$$

$$\times j_{2n-1}(2\pi Z) \left( \left( \sum_{m=1}^{n} (-1)^m \tilde{P}_{2n-1}^{2m-1}(\cos\beta) \mathbb{E}[\tilde{P}_{2n-1}^{2m-1}(\cos\theta) g_{E,V}(\theta)] \mathbb{E}[\sin((2m-1)\phi) g_{E,H}(\phi)] \right) \right.$$

$$\left. - \left( \sum_{m=1}^{n} (-1)^m \tilde{P}_{2n-1}^{2m}(\cos\beta) \mathbb{E}[\tilde{P}_{2n-1}^{2m}(\cos\theta) g_{E,V}(\theta)] \mathbb{E}[\cos(2m\phi) g_{E,H}(\phi)] \right) \right). \tag{16}$$

$$\rho_E(s-s', z-z') = \pi^2 a_\phi(0) b_\theta(1) j_0(2\pi Z) + \sum_{n=1}^{\infty} (-1)^n (4n+1) j_{2n}(2\pi Z) P_{2n}(\cos\beta) \pi^2 a_\phi(0)$$

$$\times \sum_{k=-n}^{n} p_{n-k} p_{n+k} \frac{1}{2} [b_\theta(2k+1) - b_\theta(2k-1)] - \sum_{n=1}^{\infty} i(-1)^n (4n-1) j_{2n-1}(2\pi Z) P_{2n-1}(\cos\beta) \pi^2 a_\phi(0)$$

$$\times 2 \sum_{k=1}^{n} p_{n-k} p_{n+k-1} \frac{1}{2} [b_\theta(2k) - b_\theta(2k-2)] + \sum_{n=1}^{\infty} 4(-1)^n j_{2n}(2\pi Z) \left[ \left( \sum_{m=1}^{n} (-1)^m \tilde{P}_{2n}^{2m}(\cos\beta) \frac{\pi^2}{2} a_\phi(2m) \right.\right.$$

$$\times \sum_{k=0}^{n} c_{2n,2k}^{2m} [b_\theta(2k+1) - b_\theta(2k-1)] \right) - \left( \sum_{m=1}^{n} (-1)^m \tilde{P}_{2n}^{2m-1}(\cos\beta) \frac{\pi^2}{2} b_\phi(2m-1) \sum_{k=1}^{n} d_{2n,2k}^{2m-1} [a_\theta(2k-1)$$

$$\left.\left. - a_\theta(2k+1)] \right) \right] + \sum_{n=1}^{\infty} 4i(-1)^n j_{2n-1}(2\pi Z) \left[ \left( \sum_{m=1}^{n} (-1)^m \tilde{P}_{2n-1}^{2m-1}(\cos\beta) \frac{\pi^2}{2} b_\phi(2m-1) \sum_{k=1}^{n} d_{2n-1,2k-1}^{2m-1} \right.\right.$$

$$\left.\left. \times [a_\theta(2k-2) - a_\theta(2k)] \right) - \left( \sum_{m=1}^{n} (-1)^m \tilde{P}_{2n-1}^{2m}(\cos\beta) \frac{\pi^2}{2} a_\phi(2m) \sum_{k=1}^{n} c_{2n-1,2k-1}^{2m} [b_\theta(2k) - b_\theta(2k-2)] \right) \right], \tag{24}$$

the overall antenna ports in terms of the correlation between the underlying physical elements constituting the ports and the corresponding downtilt weights.

From (10) it is evident that the SCF for the channels constituted by any two antenna ports, $s$ and $s'$, will be a function of the correlations between all the elements constituting these ports and the weight functions applied to these elements as,

$$\rho(s, s') = \mathbb{E}[[\mathbf{h}]_s [\mathbf{h}]_{s'}^H] = \sum_{z=1}^{N_E} \sum_{z'=1}^{N_E} w_s^z w_{s'}^{z'*} \mathbb{E}[g_E(\phi, \theta)$$

$$\times v_s^z(\phi, \theta) v_{s'}^{z'*}(\phi, \theta)],$$

$$= \sum_{z=1}^{N_E} \sum_{z'=1}^{N_E} w_s^z w_{s'}^{z'*} \rho_E(s-s', z-z'), \tag{28}$$

for $s, s' = 1, \ldots, N_{BS}$, where $\rho_E(s-s', z-z')$ is given by (24). The $N_{BS} \times N_{BS}$ correlation matrix for the antenna ports constituting the AAS can therefore be written as,

$$\mathbf{R}_{BS} = \tilde{\mathbf{W}}^H \mathbf{R}^E \tilde{\mathbf{W}}, \tag{29}$$

where $\tilde{\mathbf{W}}$ is a $N_{BS} N_E \times N_{BS}$ block diagonal matrix of the weight vectors applied to the $N_{BS}$ antenna ports given by,

$$\tilde{\mathbf{W}}^H = \begin{bmatrix} \mathbf{w}_1^H & \mathbf{0}^{1 \times N_E} & & \mathbf{0}^{1 \times N_E(N_{BS}-2)} \\ \mathbf{0}^{1 \times N_E} & \mathbf{w}_2^H & & \mathbf{0}^{1 \times N_E(N_{BS}-2)} \\ & & \ddots & \\ \mathbf{0}^{1 \times N_E} & \mathbf{0}^{1 \times N_E(N_{BS}-2)} & & \mathbf{w}_{N_{BS}}^H \end{bmatrix}, \tag{30}$$

where $\mathbf{w}_s$ is the $N_E \times 1$ weight vector for the $s^{th}$ antenna port and $\mathbf{R}^E$ is the $N_{BS} N_E \times N_{BS} N_E$ correlation matrix for all the elements constituting the AAS defined as,

$$[\mathbf{R}^E]_{(s'-1)N_E + z', (s-1)N_E + z} = \rho_E(s-s', z-z'), \tag{31}$$

for $z, z' = 1, \ldots, N_E$, $s, s' = 1, \ldots, N_{BS}$, where $\rho_E(s-s', z-z')$ is given by (24). With this formulation, $[\mathbf{R}_{BS}]_{s', s} = \rho(s, s')$.

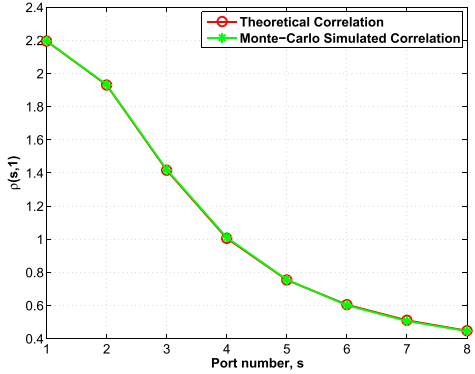

Fig. 3. Validation of the proposed SCF.

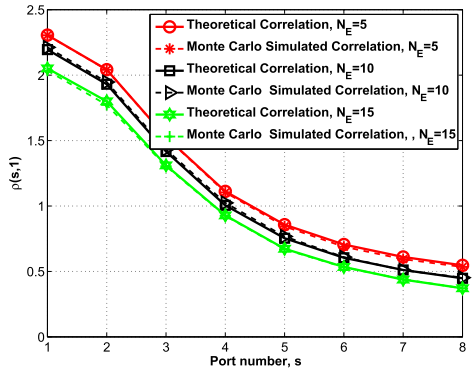

Fig. 4. Effect of $N_E$ on correlation.

In order to validate the proposed SCF, all antenna ports are assumed to transmit at a downtilt angle of $\theta_{tilt_s} = 90^o$, $s = 1, \ldots, N_{BS}$. The elevation angles are generated according to Laplacian density spectrum, with mean AoD $\theta_0$ and spread $\sigma_t$. The azimuth angles are generated using Von Mises distribution, with mean $\mu$ and spread $\propto 1/\kappa_t$. The parameter values are set as $N_0 = 30$, $\sigma_t = 15^o$, $\theta_0 = 100^o$, $\kappa_t = 10$ and $\mu = \pi/3$. The validation of the theoretical result in (28), where $\rho_E(s - s', z - z')$ is computed using (24), is done by comparison with the Monte-Carlo simulated correlation. The Monte Carlo simulations are performed over 100000 realizations of (12) to obtain the simulated $\rho_E(s - s', z - z')$. The results are shown in Fig. 3 for $N_E = 10$, $N_{BS} = 8$ and $d_y = d_z = 0.5\lambda$. The derived theoretical result provides a perfect fit to the Monte Carlo simulated correlation for only 30 summations over $n$. In Fig. 4, the correlation is seen to decrease with increasing $N_E$. The reason is attributed to the fact that for higher $N_E$, the main lobe of the antenna port radiation pattern is narrower. As a consequence, the energy of a higher number of propagation paths is not captured by the Tx beam, resulting in a decrease in the Tx power and correlation. Increasing $d_z$ will also reduce the correlation for similar reasons. The correlation between antenna ports therefore depends on the underlying arrangement of the antenna elements constituting these ports, i.e. the values of $N_E$, $d_z$ and downtilt weights.

## IV. PERFORMANCE OPTIMIZATION OF FD-MIMO SYSTEMS

The previous section expressed the spatial correlation between the antenna ports as a function of the correlations between the elements constituting the ports and the downtilt weight vectors applied to these ports. In this section, we optimize the downtilt weight vectors to maximize the downlink SNR and SIR in the single user and multi-user MISO settings respectively. Existing works on 3D beamforming optimize the downtilt angle utilizing the approximate antenna port radiation pattern expressions from ITU report [15], [26], [37]. This work makes use of the exact radiation pattern expressions that depend on the values of $N_E$, $d_z$, and the downtilt weights, and proposes solutions for the downtilt weight optimization problem. From here on, the focus is on FD-MISO settings and Rayleigh correlated channel coefficients given as follows.

**Assumption A-1.** *The channel vector, $\mathbf{h} \in \mathbb{C}^{N_{BS} \times 1}$, for a user equipped with a single isotropic Rx antenna element and served by a $N_E \times N_{BS}$ AAS deployed at the BS is modeled as,*

$$\mathbf{h} = \mathbf{R}_{BS}^{\frac{1}{2}}\mathbf{z}, \tag{32}$$

*where $\mathbf{z}$ has i.i.d zero mean, unit variance complex Gaussian entries and $\mathbf{R}_{BS}$ is the user's channel correlation matrix given by $\mathbf{R}_{BS} = \widetilde{\mathbf{W}}^H \mathbf{R}^E \widetilde{\mathbf{W}}$, where $\widetilde{\mathbf{W}}$ and $\mathbf{R}^E$ are defined in (30) and (31) respectively. The channel correlation matrix satisfies the condition,*

$$\limsup_{N_{BS}} ||\mathbf{R}_{BS}|| < +\infty. \tag{33}$$

Optimizing the downtilt weights directly using the ray-tracing model in (10) is a difficult task, given the explicit dependence of the channel on the number of paths and associated small-scale parameters (AoDs, powers). The model outlined in **Assumption A-1** will allow us to exploit tools from random matrix theory (RMT) to propose solutions for the optimal downtilt weight vectors. Moreover, it was shown in [33] that both the parametric channel model in (10) and the correlation based model in (32) give similar results in rich scattering environments.

The optimization of the downtilt angles $\theta_{tilt_s}$ of the $s = 1, \ldots, N_{BS}$ antenna ports using the 3GPP proposed expression for the weight functions in (9) is not tractable, because of the non-linear relationship between $\theta_{tilt_s}$ and $w_s^z$. Instead, this work exploits the linear dependence of the channel in (11) on the downtilt weight vector $\mathbf{w}_s$ and optimizes the elements of this weight vector directly, allowing them to take arbitrary unequal values. The proposed algorithms in the subsequent sections of this work will not yield a similar structure for the downtilt weight vectors as the one given in (9) but will result in elevation domain beams that yield significant performance gains when compared to existing elevation beamforming methods in literature.

Before formulating the optimization problems for the downtilt weight vectors, an important trace lemma is recalled that plays a key role in the problem formulations and analysis.

**Lemma 2 ([38, Lemma 14.2]).** *Let $\mathbf{A} \in \mathbb{C}^{N \times N}$ and $\mathbf{x} = [x_1, \ldots, x_N]^T \in \mathbb{C}^{N \times 1}$ be a random vector of i.i.d. entries independent of $\mathbf{A}$, such that $\mathbb{E}[x_i] = 0$, $\mathbb{E}[|x_i|^2] = 1$, $\mathbb{E}[|x_i|^8] < \infty$, and $\limsup_N ||\mathbf{A}|| < \infty$. Then,*

$$\frac{1}{N}\mathbf{x}^H\mathbf{A}\mathbf{x} - \frac{1}{N}tr\ \mathbf{A} \xrightarrow[N \to \infty]{a.s.} 0. \tag{34}$$

## A. Elevation Beamforming in a Single-User MISO System

The downlink of a single-cell MISO system is considered first, where a $N_E \times N_{BS}$ AAS at the BS serves a single user equipped with a single Rx antenna element. The received complex baseband signal $y$ at the user is given by,

$$y = \sqrt{\varrho} \mathbf{h}^H \mathbf{x} + n, \qquad (35)$$

where $\mathbf{x} \in \mathbb{C}^{N_{BS} \times 1}$ is the Tx signal from the AAS, $\mathbf{h}^H \in \mathbb{C}^{1 \times N_{BS}}$ is the channel vector from the BS to the user given by (32) and $n \sim \mathcal{CN}(0, \sigma_n^2)$ is the additive white Gaussian noise (AWGN) with variance $\sigma_n^2$ at the user. Also $\varrho$ is given as,

$$\varrho = P_{Tx} \times \text{PL} \times \text{SF} \times 10^{\frac{G_{E,max}}{10}}, \qquad (36)$$

where PL is the path loss experienced by the user, SF is the shadow fading and $P_{Tx}$ is the transmitted power. The downlink SNR for the user is then given as,

$$\gamma = \frac{\varrho}{\sigma_n^2} tr(\mathbf{h}\mathbf{h}^H). \qquad (37)$$

We are interested in finding the SNR maximizing $N_E \times 1$ weight vectors $\mathbf{w}_s$, $s = 1, \ldots, N_{BS}$, that form the $\widetilde{\mathbf{W}}$ matrix defined in (30). The optimization problem is formulated as follows.

*Problem (P1):*

$$\underset{\mathbf{w}_1, \mathbf{w}_2, \ldots, \mathbf{w}_{N_{BS}}}{\text{maximize}} \quad \gamma \qquad (38)$$

$$\text{subject to} \quad ||\mathbf{w}_s||_2 = 1, \text{ for } s = 1, \ldots, N_{BS}. \qquad (39)$$

The constraint in (39) ensures that the total power of every antenna port is bounded and does not grow indefinitely with the number of elements stacked in a port. This problem has a simple solution in the large $(N_{BS}, N_E)$ regime given in the following theorem.

*Theorem 2.* Consider a single user system consisting of a BS equipped with a $N_E \times N_{BS}$ AAS serving a single-antenna user with the channel covariance matrix $\mathbf{R}_{BS}$. Then in the large $(N_{BS}, N_E)$ regime, the optimal 3D beamforming weight vectors $\mathbf{w}_s^*$ can be computed as,

$$\mathbf{w}_s^* = \mathbf{v}_{\lambda_{max}(\mathbf{R}_{ss}^E)}, \quad s = 1, \ldots, N_{BS}, \qquad (40)$$

where $\mathbf{v}_{\lambda_{max}(\mathbf{R}_{ss}^E)}$ is the eigenvector corresponding to the maximum eigenvalue $\lambda_{max}$ of $\mathbf{R}_{ss}^E$, where $\mathbf{R}_{ss}^E$ is a $N_E \times N_E$ matrix given by $\mathbf{R}^E([(s-1)N_E + 1 : sN_E], [(s-1)N_E + 1 : sN_E])$, for $\mathbf{R}^E$ defined in (31), such that,

$$\mathbf{R}_{ss}^{E^{\frac{1}{2}}} \mathbf{w}_s^* = \sqrt{\lambda_{max}(\mathbf{R}_{ss}^E)} \mathbf{w}_s^*. \qquad (41)$$

Physically, $\mathbf{R}_{ss}^E$ is the correlation matrix formed by the elements of port $s$ and is therefore given by the $s^{th}$ $N_E \times N_E$ diagonal matrix of $\mathbf{R}^E$. The proof is postponed to Appendix A.

## B. Elevation Beamforming in a Multi-User MISO System

The downlink of a multi-user MISO system is considered next, where a $N_E \times N_{BS}$ AAS serves $K$ non-cooperating single-element users. The BS uses linear precoding in the digital domain to mitigate inter-user interference. The precoding vector and the data symbol for user $k$ are denoted as $\mathbf{g}_k \in \mathbb{C}^{N_{BS} \times 1}$ and $s_k \sim \mathcal{CN}(0, 1)$ respectively. The BS transmits the $N_{BS} \times 1$ signal,

$$\mathbf{x} = \sum_{k=1}^{K} \mathbf{g}_k s_k = \mathbf{Gs}, \qquad (42)$$

where $\mathbf{G}$ is the $N_{BS} \times K$ precoding matrix and $\mathbf{s}$ is the $K \times 1$ vector of data symbols. The received complex baseband signal at the user $k$, $y_k$, is given by,

$$y_k = \sum_{l=1}^{K} \sqrt{\varrho_k} \mathbf{h}_k^H \mathbf{g}_l s_l + n_k, \qquad (43)$$

where $\mathbf{h}_k^H \in \mathbb{C}^{1 \times N_{BS}}$ is the channel vector from the BS to the user $k$ defined using (32) as $\mathbf{h}_k = \mathbf{R}_{BS_k}^{\frac{1}{2}} \mathbf{z}_k$.

Every per user channel correlation matrix $\mathbf{R}_{BS_k}$ is given by $\mathbf{R}_{BS_k} = \widetilde{\mathbf{W}}^H \mathbf{R}_k^E \widetilde{\mathbf{W}}$ as defined in (29), where $\widetilde{\mathbf{W}}$ is the weight matrix defined in (30) and $\mathbf{R}_k^E$ is the user $k$'s elements correlation matrix with each entry given as,

$$[\mathbf{R}_k^E]_{(s'-1)N_E+z',(s-1)N_E+z} = \rho_{E_k}(s - s', z - z'), \qquad (44)$$

for $z, z' = 1, \ldots, N_E$, $s, s' = 1, \ldots, N_{BS}$, where $\rho_{E_k}(s - s', z - z')$ is computed using (24) for user $k$. Also $n_k \sim \mathcal{CN}(0, \sigma_n^2)$ is the AWGN with variance $\sigma_n^2$ at user $k$ and $\varrho_k$ is computed using (36) for user $k$.

*Remark 2.* The computation of $\rho_{E_k}(s - s', z - z')$ requires the computation of FS coefficients of the power spectra for user $k$. The FS coefficients of the PES for user $k$, $(a_{\theta_k}, b_{\theta_k})$ computed using any popular distribution, will depend only on the mean elevation AoD/line of sight (LoS) angle of user $k$ and the elevation angular spread at the BS. Similarly, the FS coefficients of the PAS for user $k$ $(a_{\phi_k}, b_{\phi_k})$ will depend only on the mean azimuth AoD/LoS angle of user $k$ and the azimuth angular spread at the BS. Therefore the correlation matrices vary across the users through their mean elevation and azimuth AoDs, which are used to determine the FS coefficients, $a_{\theta_k}, b_{\theta_k}, a_{\theta_k}$ and $b_{\theta_k}$, for $k = 1, \ldots, K$, that are then plugged into (24).

Linear precoding schemes are generally asymptotically optimal in the large $(N_{BS}, K)$ regime and robust to CSI imperfections [2], [3]. However, the complexity of computing these schemes is prohibitively high in the large $(N_{BS}, K)$ regime. A notable exception is maximum ratio transmission (MRT) [39], which is a popular scheme for large scale MIMO systems due to its low computational complexity, robustness, and high asymptotic performance [2]. Therefore, we focus on MRT precoding given by the conjugate of the channel vector $\mathbf{h}_k^H$ as,

$$\mathbf{g}_k = \beta \mathbf{h}_k, \qquad (45)$$

where $\beta$ is chosen to satisfy the Tx power constraint $tr(\mathbf{GG}^H) = 1$ as,

$$\beta = \frac{1}{\sqrt{tr(\mathbf{HH}^H)}}, \qquad (46)$$

where $\mathbf{H} = [\mathbf{h}_1 \mathbf{h}_2 \ldots \mathbf{h}_K]$.

The SINR and rate of user $k$ are defined respectively as,

$$SINR_k = \frac{\mathbf{h}_k^H \mathbf{g}_k \mathbf{g}_k^H \mathbf{h}_k}{\sum_{l \neq k}^K \mathbf{h}_k^H \mathbf{g}_l \mathbf{g}_l^H \mathbf{h}_k + \frac{\sigma^2}{\rho_k}}, \tag{47}$$

$$r_k = \log_2(1 + SINR_k) \tag{48}$$

With MRT precoding, the SINR can be written as,

$$SINR_k = \frac{|\mathbf{h}_k^H \mathbf{h}_k|^2}{\sum_{l \neq k}^K \mathbf{h}_k^H \mathbf{h}_l \mathbf{h}_l^H \mathbf{h}_k + \frac{\sigma^2}{\beta^2 \rho_k}}, \tag{49}$$

An asymptotic analysis of this quantity would yield a deterministic approximation for the $SINR$ in the large $(N_{BS}, N_E, K)$ regime as stated in the following proposition.

**Proposition 1.** Consider a multi-user MISO system consisting of a BS equipped with a $N_E \times N_{BS}$ AAS serving $K$ non-cooperating users with channel correlation matrices $\mathbf{R}_{BSk}$, $k = 1, \ldots, K$, that satisfy the condition in **Assumption A-1**. Then in the large $(N_{BS}, N_E, K)$ regime, the $SINR$ converges as (50), as shown at the bottom of the next page.

The proof of **Proposition 1** is provided in Appendix B.

Now given that $\mathbf{R}_{BSk} = \widetilde{\mathbf{W}}^H \mathbf{R}_k^E \widetilde{\mathbf{W}}$ as defined in (29), (30), and (44), the SINR converges in the large $(N_{BS}, N_E, K)$ as (51), as shown at the bottom of the next page, where $\mathbf{w}_s$ and $\mathbf{w}_{s'}$ are the $N_E \times 1$ weight vectors for the antenna ports $s$ and $s'$ that form the block diagonal matrix $\widetilde{\mathbf{W}}$ in (30), $\mathbf{R}_{k,ss}^E$ is a $N_E \times N_E$ matrix given by $\mathbf{R}_k^E([(s-1)N_E + 1 : sN_E], [(s-1)N_E + 1 : sN_E])$, where $\mathbf{R}_k^E$ is defined in (44). Similarly $\mathbf{R}_{k,ss'}^E$ is a $N_E \times N_E$ matrix given by $\mathbf{R}_k^E([(s-1)N_E + 1 : sN_E], [(s'-1)N_E + 1 : s'N_E])$.

Note that $\mathbf{R}_{k,ss'}^E$ refers to the cross-correlation matrix between the elements of port $s$ and port $s'$, given by the $(s, s')^{th}$ $N_E \times N_E$ block matrix of $\mathbf{R}_k^E$.

*1) Problem Formulation:* The focus of this section is on interference limited systems, so the performance metric employed is the max min SIR, which provides a good balance between system throughput, user fairness, and computational complexity. The deterministic approximation for SIR using (51) is given as,

$$SIR_k - \frac{(\frac{1}{N_{BS}} \sum_{s=1}^{N_{BS}} \mathbf{w}_s^H \mathbf{R}_{k,ss}^E \mathbf{w}_s)^2}{\frac{1}{N_{BS}^2} \sum_{l \neq k}^K \sum_{s,s'=1}^{N_{BS}} \mathbf{w}_s^H \mathbf{R}_{k,ss'}^E \mathbf{w}_{s'} \mathbf{w}_{s'}^H \mathbf{R}_{l,s's}^E \mathbf{w}_s} \xrightarrow{a.s.} 0, \tag{52}$$

The optimization problem is formulated as, *Problem (P2):*

$$\underset{\mathbf{w}_1, \mathbf{w}_2, \ldots \mathbf{w}_{N_{BS}}}{\text{maximize}} \quad \underset{k \in \{1, \ldots, K\}}{\text{minimize}}$$

$$\frac{(\sum_{s=1}^{N_{BS}} \mathbf{w}_s^H \mathbf{R}_{k,ss}^E \mathbf{w}_s)^2}{\sum_{l \neq k}^K \sum_{s,s'=1}^{N_{BS}} \mathbf{w}_s^H \mathbf{R}_{k,ss'}^E \mathbf{w}_{s'} \mathbf{w}_{s'}^H \mathbf{R}_{l,s's}^E \mathbf{w}_s} \tag{53}$$

$$\text{subject to } ||\mathbf{w}_s||_2 = 1, \quad s \in \{1, \ldots, N_{BS}\}. \tag{54}$$

The problem of optimizing the beamforming weights through joint max-min problem formulations has been considered in [40], where the problem of multi-cast beamforming with different receiver groups was shown to be NP-hard and was solved quasi-optimally using semi-definite relaxation (SDR). The problem at hand is much harder because it considers the joint optimization of different weight vectors for different antenna ports. Even after applying SDR and substituting the positive semi-definite rank-one matrix $\mathbf{w}_s \mathbf{w}_s^H \in \mathbb{C}^{N_E \times N_E}$ for a positive semi-definite matrix $\mathbf{W}_s \in \mathbb{C}^{N_E \times N_E}$ of arbitrary rank, the relaxed problem is not tractable because of the product of $\mathbf{W}_s$ and $\mathbf{W}_{s'}$ in the denominator of the objective function. In order to enable a tractable relaxation, all the antenna ports are assumed to transmit using the same optimal downtilt weight vector. Dropping the subscripts $s$ and $s'$, the resulting problem is given as,

*Problem (P3):*

$$\underset{\mathbf{w}}{\max} \quad \underset{k \in \{1, \ldots, K\}}{\min} \quad \frac{(\sum_{s=1}^{N_{BS}} tr(\mathbf{w}\mathbf{w}^H \mathbf{R}_{k,ss}^E))^2}{\sum_{l \neq k}^K \sum_{s,s'=1}^{N_{BS}} tr(\mathbf{w}\mathbf{w}^H \mathbf{R}_{k,ss'}^E \mathbf{w}\mathbf{w}^H \mathbf{R}_{l,s's}^E)} \tag{55}$$

$$\text{subject to } tr(\mathbf{w}\mathbf{w}^H) = 1. \tag{56}$$

This is a basic elevation beamforming scenario referred to as single downtilt beamforming (SDB). Now substituting the positive semi-definite rank-one matrix $\mathbf{w}\mathbf{w}^H \in \mathbb{C}^{N_E \times N_E}$ in *Problem (P3)* for a positive semi-definite matrix $\mathbf{W} \in \mathbb{C}^{N_E \times N_E}$ of arbitrary rank, the semi-definite relaxed problem is given as,

*Problem (P4):*

$$\underset{\mathbf{W}}{\max} \quad \underset{k \in \{1, \ldots, K\}}{\min} \quad \frac{(\sum_{s=1}^{N_{BS}} tr(\mathbf{W}\mathbf{R}_{k,ss}^E))^2}{\sum_{l \neq k}^K \sum_{s,s'=1}^{N_{BS}} tr(\mathbf{W}\mathbf{R}_{k,ss'}^E \mathbf{W}\mathbf{R}_{l,s's}^E)} \tag{57}$$

$$\text{subject to } \mathbf{W} \succeq 0, tr(\mathbf{W}) = 1. \tag{58}$$

*Problem (P4)* is efficiently solved using fractional programmings tools as discussed now.

*2) Optimization Technique and Solution:* Fractional programming provides efficient tools to maximize the minimum of ratios in which the numerator is a concave function, the denominator is a convex function, and the constraint set is convex, whereas no low-complexity optimization method is available if any of these properties is not met [41], [42]. An efficient method to do so is the generalized Dinkelbach's algorithm, discussed in Appendix A of [42]. In order to meet the conditions for application of Dinkelbach's method, *Problem (P4)* is reformulated as,

*Problem (P5):*

$$\underset{\mathbf{W}}{\max} \quad \underset{k \in \{1, \ldots, K\}}{\min} \quad \frac{\sum_{s=1}^{N_{BS}} tr(\mathbf{W}\mathbf{R}_{k,ss}^E)}{\sqrt{\sum_{l \neq k}^K \sum_{s,s'=1}^{N_{BS}} tr(\mathbf{W}\mathbf{R}_{k,ss'}^E \mathbf{W}\mathbf{R}_{l,s's}^E)}} \tag{59}$$

$$\text{subject to } \mathbf{W} \succeq 0, tr(\mathbf{W}) = 1. \tag{60}$$

The objective function in (59) considers a set of ratios of two functions, where we denote the numerator by $f_k(\mathbf{W})$ and the denominator by $g_k(\mathbf{W})$, $k = 1, \ldots, K$. In order to study these functions, the following properties of the $vec$ function are exploited.

**Lemma 3.** For any matrix $\mathbf{A} \in \mathbb{C}^{M \times N}$, the $vec$ operator is defined as [43],

$$vec(\mathbf{A}) = (a_{11}, \ldots, a_{M1}, a_{12}, \ldots, a_{M2}, \ldots, a_{1N}, \ldots, a_{MN})^T. \tag{61}$$

Some properties of the $vec$ operator are:

$$tr(\mathbf{AB}) = vec(\mathbf{A}^T)^T vec(\mathbf{B}), \quad \forall \mathbf{A}, \mathbf{B} \in \mathbb{C}^{M \times M}, \tag{62}$$

**Algorithm 1** Single Downtilt Beamforming (SDB)

1: **procedure** GENERALIZED DINKELBACH($\mathbf{W}$)
2:  Set $\epsilon > 0$;
3:  Initialize $\lambda = 0$;
4:  **repeat**
5:   $\mathbf{W}^* = \max\limits_{\mathbf{W} \in \mathbb{C}^{N_E \times N_E}} \{ \min\limits_{1 \le k \le K} [f_k(\mathbf{W}) - \lambda g_k(\mathbf{W})] \}$,
      where $f_k(\mathbf{W})$ and $g_k(\mathbf{W})$ are given by
      (64) and (66) respectively, subject to
      $\mathbf{W} \succeq 0$ and $tr(\mathbf{W}) = 1$;
6:   $F = \min\limits_{1 \le k \le K} \{ f_k(\mathbf{W}^*) - \lambda g_k(\mathbf{W}^*) \}$;
7:   $\lambda = \min\limits_{1 \le k \le K} f_k(\mathbf{W}^*)/g_k(\mathbf{W}^*)$;
8:  **until** $F < \epsilon$.
9: **procedure** GAUSSIAN RANDOMIZATION($\mathbf{w}$)
10:  **for** $l = 1$ to $L$
11:   Generate $\boldsymbol{\zeta}_l \sim \mathcal{CN}(\mathbf{0}, \mathbf{W}^*)$;
12:   Construct a feasible solution $\mathbf{w}_l = \text{sgn}(\zeta_l)/\sqrt{N_E}$;
13:  **end for**
14:  Determine $l^* = \max\limits_{l=1,...,L} \min\limits_{1 \le k \le K} SIR_k(\mathbf{w}_l)$, where $SIR_k$
      is given by (55);
15:  $\mathbf{w}^* = \mathbf{w}_{l^*}$.

$$tr(\mathbf{A}^T \mathbf{BCD}^T) = vec(\mathbf{A})^T (\mathbf{D} \otimes \mathbf{B}) vec(\mathbf{C}), \qquad (63)$$
$$\forall \mathbf{A}, \mathbf{B}, \mathbf{C}, \mathbf{D} \in \mathbb{C}^{M \times M}.$$

Exploiting these properties, $f_k(\mathbf{W})$ and $g_k(\mathbf{W})$ can be expressed as,

$$f_k(\mathbf{W}) = \sum_{s=1}^{N_{BS}} vec(\mathbf{W}^T)^T vec(\mathbf{R}_{k,ss}^E). \qquad (64)$$

$$g_k(\mathbf{W}) = \sqrt{\sum_{l \ne k}^{K} \sum_{s,s'=1}^{N_{BS}} vec(\mathbf{W}^T)^T (\mathbf{R}_{l,s's}^{E}{}^T \otimes \mathbf{R}_{k,ss'}^E) vec(\mathbf{W})} \qquad (65)$$

$$= \left|\left| \left( \sum_{l \ne k}^{K} \sum_{s,s'=1}^{N_{BS}} (\mathbf{R}_{l,s's}^{E}{}^T \otimes \mathbf{R}_{k,ss'}^E) \right)^{\frac{1}{2}} vec(\mathbf{W}) \right|\right|_2. \qquad (66)$$

It can be seen from (64) that $f_k(\mathbf{W})$ is a linear function of $\mathbf{W}$. Also $g_k(\mathbf{W})$ is a convex function, expressed as an L2 norm in (66). *Problem (P5)* therefore considers a set of ratios $\{\frac{f_k(\mathbf{W})}{g_k(\mathbf{W})}\}_{k=1}^{K}$, where each ratio has an affine numerator $f_k(\mathbf{W})$, convex denominator $g_k(\mathbf{W})$ and convex constraints and can therefore be solved using the generalized Dinkelbach's algorithm [42]. The Dinkelbach's procedure to solve *Problem (P5)* is formulated in **Algorithm 1**. Once the optimal $\mathbf{W}^*$ is obtained, the corresponding weight vector $\mathbf{w}$ that solves the *Problem (P3)* needs to be extracted. Generally the resulting matrix $\mathbf{W}^*$, although globally optimal for *Problem (P5)* has a rank greater than one and therefore yields a quasi-optimal solution for *Problem (P3)*. It is important to post-process the relaxed solution $\mathbf{W}^*$ to extract a close to optimal $\mathbf{w}^*$. Besides the eigenvector approximation method, where $\mathbf{w}^*$ is approximated as the principal eigenvector of $\mathbf{W}^*$, randomization is another way to extract an approximate solution from the SDR solution $\mathbf{W}^*$. The idea is to generate a random vector $\boldsymbol{\zeta} \in \mathbb{C}^{N_E \times 1} \sim \mathcal{CN}(\mathbf{0}, \mathbf{W}^*)$ and use it to construct an approximate solution to *Problem (P3)*. The procedure and theoretical accuracy results have been discussed in [40], [44]. The specific design of the randomization procedure is problem-dependent and has been summarized at the end of **Algorithm 1**.

The proposed SDB algorithm splits downlink beamforming into two linear stages: an elevation beamforming stage that depends only on the users' channel correlation matrices to maximize the minimum user SIR of the system and the MRT precoding stage for the effective channels with dimension reduced from $N_{BS}N_E$ to $N_{BS}$. This channel dimension reduction can be seen in (10), where the channel with antenna port $s$ is a weighted sum of the channels with the $N_E$ elements inside port $s$. The feedback overhead for the implementation of the digital precoding stage is therefore significantly reduced, since the number of antenna ports is much less than the total number of antenna elements.

The channel correlation matrix for each user can be computed at the BS exploiting the theoretical expression in (24), which requires knowledge of only the angular spread at the BS (locally estimated) and the user's LoS azimuth and elevation angles. The BS can estimate the location of each user in the uplink and compute the corresponding LoS angle. In this work, we have assumed that the BS has perfect knowledge of the users' channel correlation matrices, which can be accurately learned and tracked since they are constant in time. Even for nomadic users, the correlation matrices evolve in time much more slowly than the actual Rayleigh fading process as discussed in [21] and can be tracked using well-known existing algorithms.

## V. USER GROUP SPECIFIC ELEVATION BEAMFORMING

The last section focused on the SDB scenario, where all the users are served by vertical beams transmitted using the same optimal downtilt antenna port weight vector.

### A. Motivation

In order to minimize the loss incurred by the single downtilt assumption, we further propose the user-group specific single downtilt beamforming (UG-SDB) scenario, where the user population is partitioned into groups based on their channel

$$SINR_k - \frac{\frac{1}{N_{BS}^2}( tr\mathbf{R}_{BSk})^2}{\frac{1}{N_{BS}^2}(\sum_{l \ne k}^{K} tr(\mathbf{R}_{BSk}\mathbf{R}_{BSl}) + \sum_{k=1}^{K} tr(\mathbf{R}_{BSk})\frac{\sigma^2}{\rho_k})} \xrightarrow{a.s.} 0, \quad k = 1, \ldots, K. \qquad (50)$$

$$SINR_k - \frac{(\frac{1}{N_{BS}}\sum_{s=1}^{N_{BS}} \mathbf{w}_s^H \mathbf{R}_{k,ss}^E \mathbf{w}_s)^2}{\frac{1}{N_{BS}^2}(\sum_{l \ne k}^{K} \sum_{s,s'=1}^{N_{BS}} \mathbf{w}_s^H \mathbf{R}_{k,ss'}^E \mathbf{w}_{s'} \mathbf{w}_{s'}^H \mathbf{R}_{l,s's}^E \mathbf{w}_s + \sum_{k=1}^{K} \sum_{s=1}^{N_{BS}} \mathbf{w}_s^H \mathbf{R}_{k,ss}^E \mathbf{w}_s \frac{\sigma^2}{\rho_k})} \xrightarrow{a.s.} 0, \qquad (51)$$

correlation matrices and each group is served by a set of antenna ports fed with the weight vector optimized for that particular user group. Different user groups are therefore served by different elevation domain beams and MRT precoding is used within each group. The grouping ensures that each downtilt weight vector is optimized using the statistics of a smaller number of co-located users, resulting in the better design of each elevation domain beam.

This idea of user grouping was effectively utilized in [21] in the context of joint spatial division and multiplexing (JSDM) approach, wherein the authors partitioned the users into groups based on two qualitative principles - first, that the users in the same group have channel covariance eigenspaces that approximately span a given common group subspace and second that different user groups have almost orthogonal subspaces. The authors proposed two schemes, the K-means clustering and the fixed quantization method, to cluster the users. The simulation results showed the superior performance of the fixed quantization algorithm, which motivated us to use the counterpart of this scheme in the FD-MIMO setting.

### B. Method

The proposed scheme considers $G$ user groups, where the group subspaces denoted by $\mathbf{V}_g \in \mathbb{C}^{M \times r_g}$; $g = 1, \ldots, G$, $M = N_{BS} \times N_E$, are fixed and known apriori based on the geometric arrangement of the users in the cell. The choice of these subspaces is critical to the performance of the proposed algorithm. The method employed proposes to group the users such that the elevation angular supports for different user groups are disjoint. Therefore, we choose $G$ mean elevation AoDs, $\theta_{0,g}$, and a fixed value for the group elevation angular spread $\triangle$, such that the resulting intervals $[\theta_{0,g} - \triangle, \theta_{0,g} + \triangle]$ are disjoint. The $M \times M$ correlation matrices $\mathbf{R}_g^E$, $g = 1, \ldots, G$, for these $G$ sets of mean elevation AoDs and angular spread are formed using (31) and the corresponding eigenspaces $\mathbf{V}_g$, $g = 1, \ldots, G$, are computed. These eigenspaces constitute the group subspaces. The $K$ users are then assigned to the $G$ groups, based on the chordal distance between the users' channel correlation eigenspaces and the group subspaces. The number of users in each group is denoted by $K_g$, $g = 1, \ldots, G$.

At the transmitter side, the antenna ports are partitioned in $G$ groups, where the number of antenna ports in each group is $N_{BS,g} = N_{BS}/G$, with $G$ chosen as a factor of $N_{BS}$. The first $N_{BS,g}$ adjacent ports in Fig. 2 serve the first user group and so on. The optimal weight vector for the $g^{th}$ antenna port group, $\mathbf{w}_g^*$, is obtained using **Algorithm 1**, utilizing the channel correlation matrices of the users in the $g^{th}$ group. Therefore $G$ different elevation beams are designed. The UG-SDB technique is summarized in **Algorithm 2** and will be shown to yield excellent performance gains in the next section in the large ($N_{BS}, N_E$) regime.

## VI. Results and Discussions

The performance gains realizable through the careful design of the downtilt antenna port weight vectors are now studied

---

**Algorithm 2** User Group Specific Single Downtilt Beamforming (UG-SDB)

1: **procedure** User Grouping($\mathcal{S}_g$)
2:    **for** $g = 1$ to $G$
3:      Initialize the user group set $\mathcal{S}_g = \emptyset$;
4:      Choose $\theta_{0,g}$ and elevation spread $\triangle$ such that the intervals $[\theta_{0,g} - \triangle, \theta_{0,g} + \triangle]$ are disjoint;
5:      Compute $\mathbf{R}_g^E(\theta_{0,g}, \triangle)$ using (44) for the chosen $\theta_{0,g}$ and $\triangle$;
6:      Obtain the group subspace $\mathbf{V}_g = \mathbf{U}_g$, where $\mathbf{U}_g$ is the $M \times r_g$ matrix of eigenvectors corresponding to $r_g$ dominant eigenvalues of $\mathbf{R}_g^E$, with $r_g$ chosen such that $\sum_{g=1}^G r_g = M$;
7:    **end for**
8:    **for** $k = 1$ to $K$
9:      Compute $\mathbf{R}_k^E$ using (44) for the propagation scenario under study;
10:     Obtain the eigenspace $\mathbf{U}_k^{M \times r_k}$, corresponding to the $r_k$ dominant eigenvalues of $\mathbf{R}_k^E$;
11:     Compute $d_C(\mathbf{U}_k, \mathbf{V}_g) = ||\mathbf{U}_k \mathbf{U}_k^H - \mathbf{V}_g \mathbf{V}_g^H||_F^2$, $g = 1, \ldots, G$;
12:     Find $g = \min_{1 \leq g' \leq G} d_C(\mathbf{U}_k, \mathbf{V}_{g'})$;
13:     Add user $k$ to group $g$, i.e. $\mathcal{S}_g := \mathcal{S}_g \cup \{k\}$;
14:    **end for**
15: **procedure** Weight Vector Optimization($\mathbf{w}_g$)
16:    Divide $N_{BS}$ antenna ports into $G$ equal groups, with $N_{BS,g}$ ports serving each user group.
17:    **for** $g = 1$ to $G$
18:     Use **Algorithm 1** to obtain the quasi-optimal weight vector $\mathbf{w}_g^*$ for the antenna ports in group $g$ serving the $\mathcal{S}_g$-user MISO system.
19:    **end for**

---

using simulations with parameter values set as $\theta_{3dB}, \phi_{3dB} = 65^o$, $\sigma_t = 15^o$, $\kappa_t = 10$, $\mu = 0$, $G_{E,max} = 8$dBi and $N_0 = 30$.

### A. Performance of **Theorem 2** in the Single User MISO Setting

The single-user MISO case is studied first, where **Theorem 2** is used to optimize the weight vectors $\mathbf{w}_s$, of the $s = 1, \ldots, N_{BS}$ antenna ports. The BS equipped with a $10 \times N_{BS}$ AAS serves an outdoor user located at the edge of a cell of radius 250m, with the LoS angle $\theta_0$ computed to be $95.37^o$. The user throughput for the optimal downtilt weight vectors is plotted in red in Fig. 5 along with the cases where the electrical downtilt angles are set to specific pre-defined values with weights computed using (9). It is evident that choosing the weight vectors according to **Theorem 2** yields higher capacity values. Also, the theoretical capacity obtained using the deterministic approximation of the SNR in (70) approximates the Monte-Carlo simulated capacity obtained using (37) quite well.

### B. Performance Comparison of SDB in the Multi-User MISO Setting

In this section, we study the multi-user MISO system with $K$ users placed randomly in a cell of radius 250m, at a minimum distance of $30m$ from the BS, where $N_{BS} > K$. Simulations results are used to confirm the performance gains of the proposed SDB algorithm by comparing it with the following downtilting strategies.

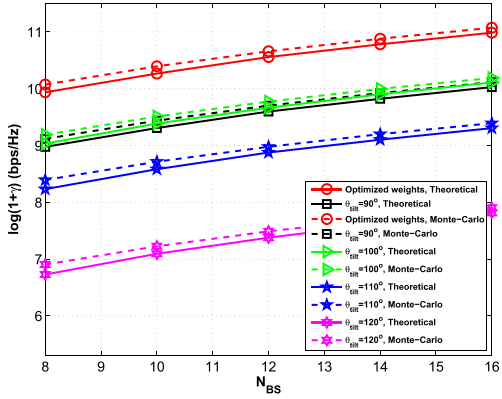

Fig. 5. Performance of a single user MISO system.

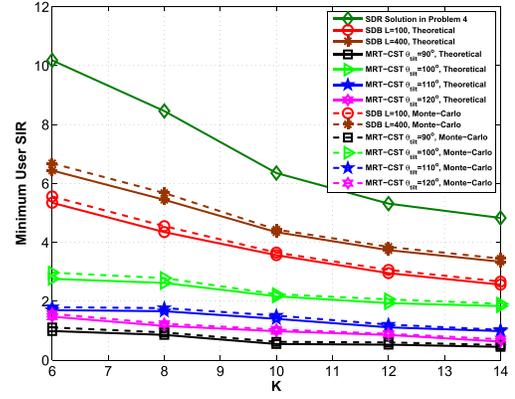

Fig. 6. Performance of SDB in a multi-user MISO system.

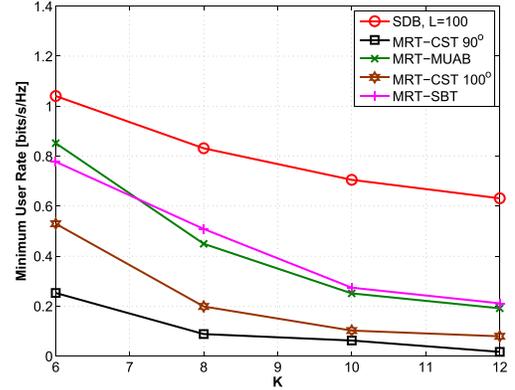

Fig. 7. Comparison of SDB with other tilting strategies for a $10 \times 12$ AAS.

- **Cell-specific tilting (CST):** The first strategy uses a common fixed tilt denoted as $\theta_{tilt}$ for all antenna ports with MRT precoding in the horizontal domain [26]. The downtilt weights corresponding to the tilt angle are computed using the 3GPP proposed weight expression given in (9).
- **Switched Beam Tilting (SBT):** In this tilting strategy proposed in [23], the cell is partitioned into two vertical regions and one of the two $(\theta_{tilt}, \theta_{3dB})$ pairs is applied at the BS when serving each region. The pairs are calculated as $(113.75^o, 21^o)$ and $(96.51^o, 6.5^o)$. The transmission using MRT precoding is scheduled to one of the two vertical regions in each time slot. The vertical region activity factor for region $s$ is given as $\frac{|\mathcal{K}_s|}{K}$, where $\mathcal{K}_s$ is the set of users in the vertical region $s$. This strategy is denoted as MRT-SBT.
- **Multi-User Active Beamforming (MUAB):** In this tilting strategy proposed in [27], the approximate antenna port radiation pattern expression given in [24] is utilized. Under the high SNR approximation, the optimal downtilt angle is computed as the weighted arithmetic mean value of the elevation LoS angles of all the users. This strategy is studied with MRT precoding and the resulting system is denoted as MRT-MUAB respectively.

The first result uses a $10 \times 40$ AAS to compare the SIR performance of SDB to that of MRT-CST, where the latter is simulated for several values of $\theta_{tilt}$ and the corresponding weights are computed using (9). The Monte-Carlo simulated minimum user $SIR$ is plotted in Fig. 6 along with the deterministic equivalent in (52). The Gaussian randomization in **Algorithm 1** is implemented using $L = 100$ and $400$ iterations. The deterministic equivalent matches the Monte-Carlo result quite well for moderate number of antennas. More importantly, the potential of elevation beamforming in enhancing the system performance is confirmed as the quasi-optimal downtilt weight vector obtained using the SDB algorithm yields significant performance gains when compared to the existing implementations with fixed downtilt angles.

The SDR solution in (57) is also plotted in Fig. 6, using the optimal positive semi-definite matrix $\mathbf{W}^*$ obtained through the Dinkelbach's method. This acts as an upper bound on the performance of the SDB algorithm, since $\mathbf{W}^*$, although a globally optimal solution for Problem (*P4*), has a rank greater than one and therefore yields a quasi-optimal solution for the weight vector in (*P3*). This quasi-optimal weight vector $\mathbf{w}^*$ is extracted from $\mathbf{W}^*$ using the Gaussian randomization technique. The loss incurred due to the semi-definite relaxation assumption can be reduced by increasing $L$ in **Algorithm 1** as seen in Fig. 6.

Fig. 7 compares the minimum user rate performance of SDB to that of the three tilting strategies outlined above for a $10 \times 12$ AAS. The proposed SDB algorithm performs much better than the other strategies. The vertical sectorization in SBT ensures that the $K_s$ users in the active region $s$ are served by a tilt angle which improves the received SNR over the active region only instead of the entire cell resulting in a better achievable minimum user rate performance at higher values of $K$ when compared with MRT-MUAB. The latter sets the downtilt as the mean of the elevation LoS angles of all the $K$ users, who might be quite scattered in the elevation.

### C. Performance of User Group Specific Single Downtilt Beamforming Scenario

We now study the performance of the proposed UG-SDB scenario where $K$ users and $N_{BS}$ antenna ports are divided into $G$ groups. The antenna ports in each group are fed with the optimal downtilt weight vector obtained using the statistics of the users being served in that group. The simulation is done for $N_{BS} = 36$, $N_E = 5$ and $G = 3$. The values for $\theta_{0,g}$, $g = 1, \ldots, G$ and $\triangle$ are set as $[93^o, 101^o, 109^o]$ and $7.5^o$ respectively. The user grouping is done as explained in **Algorithm 2**, where $r_g^* = 60$ and the optimal weight vector $\mathbf{w}_g^*$ for each antenna ports group, comprising of $N_{BS,g} = 12$ ports, is computed using the SDB **Algorithm 1**. The result for the minimum user SIR is plotted in Fig. 8, which clearly highlights the performance gains that can be achieved by user-grouping. The gap between the curves with and without grouping starts to decrease as $K$ increases, because for higher $K$, a larger number of users $K_g$ are served in each group by the same $N_{BS,g}$ antenna ports per group.

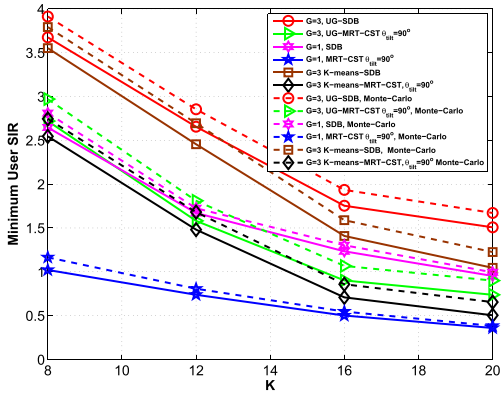

Fig. 8. Performance of user-group specific single downtilt beamforming in a multi-user MISO system.

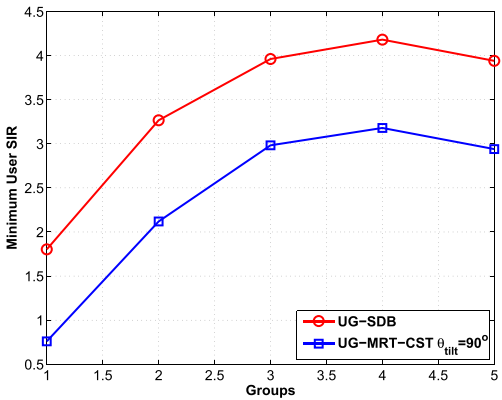

Fig. 9. Effect of the number of user groups on the performance of a multi-user MISO system.

The proposed user-grouping method is compared with K-means clustering, which is a standard iterative algorithm that partitions the observations into groups such that each observation belongs to the group with the nearest mean. Our proposed user-grouping method motivated by the fixed quantization method in [21] performs better than K-means, which is also in accordance with the observation of the authors in [21].

In the next figure, we study the effect of increasing the number of user groups. The results are plotted for $G = 2, 3, 4$ and 5 in Fig. 9, where $\theta_{0,g}$, $g = 1, \ldots, G$ and $\triangle$ are selected, ensuring that the resulting angular supports are disjoint. A $5 \times 60$ AAS is utilized for $K = 30$ users. The quasi-optimal weight vector $\mathbf{w}_g^*$, for antenna ports group $g$ serving user group $g$ is computed following the steps in **Algorithm 2**. As the number of groups increases upto four, the SIR performance improves since every user group is now served with an optimal downtilt weight vector determined using the spatial statistics of only the users in that group. A higher number of groups realizes a higher degree of vertical separation of the users through the design of a higher number of elevation domain beams, where each beam serves a small number of co-located users. However, depending on the positions of the users in the cell, it becomes more and more likely to have $K_g > N_{BS_g}$ in one of the groups as $G$ increases, requiring user scheduling in that group, while the overall number of users $K < N_{BS}$. As the ratio of $K_g$ to $N_{BS_g}$ becomes high in group $g$, the minimum SIR under grouping starts to deteriorate as compared to that with a lower number of groups.

These founding results provide a flavor of the performance gains realizable through FD-MIMO techniques. More sophisticated 3D beamforming methods can be devised in the future that allow every port to transmit at a different optimal downtilt angle. In fact, for $N_{BS} >> K$, spatially separated beams to almost all the users can be realized, which is why FD beamforming can be highly advantageous when amalgamated with massive MIMO techniques.

## VII. Conclusion

This paper reviewed the recent development of FD-MIMO technology for evolution towards 5G cellular systems and studied the architecture of the 2D active antenna arrays utilized by these systems. The 2D AAS not only serves as a practical implementation of massive MIMO systems, but also has the potential to boost spectral efficiency by providing the ability of adaptive electronic beam control in both the elevation and azimuth dimensions. The SCF for the channels constituted by individual antenna elements in the AAS was derived and used to compute the correlation between the overall antenna ports for any arbitrary 3D propagation environment. The performance benefits of FD-MIMO techniques were then studied by devising elevation beamforming algorithms, that optimize the downtilt antenna port weight vectors in the single user and multi-user MISO settings, utilizing the quasi-static channel correlation matrices of the users obtained from the derived SCF. The problem of determining the downtilt weight vector that maximizes the minimum SIR of the multi-user MISO system was formulated under the assumption that all ports serve using the same optimal downtilt weight vector, and solved using SDR and Dinkelbach's method. Finally, the user-group specific single downtilt beamforming scenario was devised. Simulation results confirmed the potential of FD-MIMO techniques to improve the system performance.

## Appendix A
### Proof of Theorem 2

This theorem follows from expressing $tr(\mathbf{h}\mathbf{h}^H)$ as a quadratic term in $\mathbf{z}$ using (32) to re-write the SNR in (37) as follows,

$$\gamma = \frac{\varrho}{\sigma_n^2} \mathbf{h}^H \mathbf{h} = \frac{\varrho}{\sigma_n^2} \mathbf{z}^H \mathbf{R}_{BS}^{\frac{1}{2}^H} \mathbf{R}_{BS}^{\frac{1}{2}} \mathbf{z}. \quad (67)$$

Next exploiting **Lemma 2** and the fact that the covariance matrix $\mathbf{R}_{BS}$ defined in (29) is Hermitian, positive semi-definite

$$\frac{1}{N_{BS}}SINR_k - \frac{(\frac{1}{N_{BS}}tr(\mathbf{R}_{BSk}))^2}{\sum_{l \neq k}^{K} \frac{1}{N_{BS}}tr(\mathbf{R}_{BSk}\mathbf{h}_l\mathbf{h}_l^H) + \frac{\sigma^2 \sum_{k=1}^{K} \frac{1}{N_{BS}}tr(\mathbf{R}_{BSk})}{\varrho_k}} \xrightarrow{a.s.} 0. \tag{75}$$

$$SINR_k - \frac{(\frac{1}{N_{BS}}tr(\mathbf{R}_{BSk}))^2}{\frac{1}{N_{BS}^2}(\sum_{l \neq k}^{K} tr(\mathbf{R}_{BSl}^{\frac{1}{2}H}\mathbf{R}_{BSk}\mathbf{R}_{BSl}^{\frac{1}{2}}) + \frac{\sigma^2 \sum_{k=1}^{K} tr(\mathbf{R}_{BSk})}{\varrho_k})} \xrightarrow{a.s.} 0. \tag{76}$$

and satisfies the condition in Assumption A-1 we have,

$$\frac{1}{N_{BS}}\mathbf{z}^H\mathbf{R}_{BS}\mathbf{z} - \frac{1}{N_{BS}}tr(\mathbf{R}_{BS}) \xrightarrow{a.s.} 0. \tag{68}$$

where, $\quad tr(\mathbf{R}_{BS}) = \sum_{s=1}^{N_{BS}} \mathbf{w}_s^H\mathbf{R}_{ss}^E\mathbf{w}_s, \tag{69}$

and $\mathbf{R}_{ss}^E$ is a $N_E \times N_E$ matrix given by $\mathbf{R}^E([(s-1)N_E + 1 : sN_E], [(s-1)N_E + 1 : sN_E])$, where $\mathbf{R}^E$ is defined in (31). Therefore in the large $(N_{BS}, N_E)$ regime,

$$\frac{1}{N_{BS}}\gamma - \frac{\varrho}{N_{BS}\sigma_n^2}\sum_{s=1}^{N_{BS}} \mathbf{w}_s^H\mathbf{R}_{ss}^E\mathbf{w}_s \xrightarrow{a.s.} 0. \tag{70}$$

Consequently, in the large $(N_{BS}, N_E)$ regime, the optimization problem $(P1)$ can be written as,

$$\underset{\mathbf{w}_1, \mathbf{w}_2, \dots, \mathbf{w}_{N_{BS}}}{\text{maximize}} \sum_{s=1}^{N_{BS}} \mathbf{w}_s^H\mathbf{R}_{ss}^E\mathbf{w}_s \tag{71}$$

$$\text{subject to} \quad ||\mathbf{w}_s||_2 = 1, \quad \text{for } s = 1, \dots, N_{BS}. \tag{72}$$

Problem $(P1)$ is equivalent to finding the optimal $\mathbf{w}_s^* = \arg\max_{\mathbf{w}} ||\mathbf{R}_{ss}^{E\frac{1}{2}}\mathbf{w}||_2^2$, $s = 1, \dots, N_{BS}$, which has the simple eigenvector solution stated in **Theorem 2**.

## APPENDIX B
### PROOF OF PROPOSITION 1

In order to prove **Proposition 1**, the following theorem will be required.

**Theorem 3.** Continuous Mapping Theorem [45]. Let $\{X_n\}$ be a sequence of $N$-dimensional random vectors and $g : \mathbb{R}^N \to \mathbb{R}^L$ be a continuous function. Then,

$$X_n \xrightarrow{a.s.} X \implies g(X_n) \xrightarrow{a.s.} g(X), \tag{73}$$

where $\xrightarrow{a.s.}$ denotes almost sure convergence.

To prove **Proposition 1**, note that the channel vector for user $k$, $\mathbf{h}_k$ is given by $\mathbf{R}_{BSk}^{\frac{1}{2}}\mathbf{z}_k$, where the condition described in **Assumption A-1** on the Hermitian, positive semi-definite channel correlation matrices $\mathbf{R}_{BSk}$ holds for all users, $k = 1, \dots, K$. With this, (49) can be written as,

$$SINR_k = \frac{|\mathbf{z}_k^H\mathbf{R}_{BSk}\mathbf{z}_k|^2}{\sum_{l \neq k}^{K} \mathbf{z}_k^H\mathbf{R}_{BSk}^{\frac{1}{2}H}\mathbf{h}_l\mathbf{h}_l^H\mathbf{R}_{BSk}^{\frac{1}{2}}\mathbf{z}_k + \frac{\sigma^2 \sum_{k=1}^{K} \mathbf{z}_k^H\mathbf{R}_{BSk}\mathbf{z}_k}{\varrho_k}}. \tag{74}$$

Applying **Lemma 2** along with the continuous mapping theorem on the numerator and **Lemma 2** along with the fact that if $\mathbf{A} \in \mathbb{C}^{M \times N}$ and $\mathbf{B} \in \mathbb{C}^{N \times M}$, then $tr(\mathbf{AB}) = tr(\mathbf{BA})$ on the denominator would yield the convergence result for (74) as shown in (75), at the top of this page.

Substituting the expression for $\mathbf{h}_l$ as $\mathbf{R}_{BSl}^{\frac{1}{2}}\mathbf{z}_l$ and applying the convergence theorem in **Lemma 2** a second time on the denominator would yield (76), as shown at the top of this page. Again using $tr(\mathbf{AB}) = tr(\mathbf{BA})$, would complete the proof of **Proposition 1**.

## REFERENCES

[1] E. Dahlman, S. Parkvall, J. Sköld, and P. Beming, *3G Evolution: HSPA and LTE for Mobile Broadband*, 2nd ed. San Diego, CA, USA: Academic, 2008.

[2] F. Rusek *et al.*, "Scaling up MIMO: Opportunities and challenges with very large arrays," *IEEE Signal Process. Mag.*, vol. 30, no. 1, pp. 40–60, Jan. 2013.

[3] T. L. Marzetta, "Noncooperative cellular wireless with unlimited numbers of base station antennas," *IEEE Trans. Wireless Commun.*, vol. 9, no. 11, pp. 3590–3600, Nov. 2010.

[4] L. Lu, G. Y. Li, A. L. Swindlehurst, A. Ashikhmin, and R. Zhang, "An overview of massive MIMO: Benefits and challenges," *IEEE J. Sel. Topics Signal Process.*, vol. 8, no. 5, pp. 742–758, Oct. 2014.

[5] Y.-H. Nam *et al.*, "Full-dimension MIMO (FD-MIMO) for next generation cellular technology," *IEEE Commun. Mag.*, vol. 51, no. 6, pp. 172–179, Jun. 2013.

[6] *Study on 3D Channel Model for LTE*, document 3GPP TR 36.873 V12.0.0, Sep. 2014.

[7] *Study of Radio Frequency (RF) and Electromagnetic Compatibility (EMC) Requirements for Active Antenna Array System (AAS) Base Station*, document 3GPP TR 37.840 V12.0.0, Mar. 2013.

[8] K. Linehan and R. Chandrasekaran, "Active antennas: The next step in radio and antenna evolution," CommScope, Hickory, NC, USA, White Paper WP-105435, 2011. [Online]. Available: https://www.yumpu.com/en/document/view/8991085/active-antennas-the-next-step-in-radio-and-antenna-evolution

[9] Huawei. (Nov. 2012). *Active Antenna System: Utilizing the Full Potential of Radio Sources in the Spatial Domain*. [Online]. Available: http://www1.huawei.com/en/static/AAS-129092-1-197969.pdf

[10] *Study on 3D-Channel Model for Elevation Beamforming and FD-MIMO Studies for LTE*, document R1-122034, 3GPP TSG RAN Plenary 58, Barcelona, Spain, Dec. 2012.

[11] *Study on Elevation Beamforming/Full-Dimension (FD) MIMO for LTE*, document 3GPP TR 36.897 V13.0.0, Jun. 2015.

[12] B. Debaillie *et al.*, "Analog/RF solutions enabling compact full-duplex radios," *IEEE J. Sel. Areas Commun.*, vol. 32, no. 9, pp. 1662–1673, Sep. 2014.

[13] Y. Song, X. Yun, S. Nagata, and L. Chen, "Investigation on elevation beamforming for future LTE-advanced," in *Proc. IEEE Int. Conf. Commun. Workshops (ICC)*, Jun. 2013, pp. 106–110.

[14] A. Kammoun, H. Khanfir, Z. Altman, M. Debbah, and M. Kamoun, "Preliminary results on 3D channel modeling: From theory to standardization," *IEEE J. Sel. Areas Commun.*, vol. 32, no. 6, pp. 1219–1229, Jun. 2014.

[15] W. Lee, S.-R. Lee, H.-B. Kong, and I. Lee, "3D beamforming designs for single user MISO systems," in *Proc. IEEE Global Commun. Conf. (GLOBECOM)*, Dec. 2013, pp. 3914–3919.

[16] Y. Kim *et al.*, "Full dimension MIMO (FD-MIMO): The next evolution of MIMO in LTE systems," *IEEE Wireless Commun.*, vol. 21, no. 2, pp. 26–33, Apr. 2014.

[17] J. Koppenborg, H. Halbauer, S. Saur, and C. Hoek, "3D beamforming trials with an active antenna array," in *Proc. ITG Workshop Smart Antennas*, 2012, pp. 110–114.

[18] F. Sohrabi and W. Yu, "Hybrid digital and analog beamforming design for large-scale antenna arrays," *IEEE J. Sel. Topics Signal Process.*, vol. 10, no. 3, pp. 501–513, Apr. 2016.

[19] R. Rajashekar and L. Hanzo, "Hybrid beamforming in mm-wave MIMO systems having a finite input alphabet," *IEEE Trans. Commun.*, vol. 64, no. 8, pp. 3337–3349, Aug. 2016.

[20] J. Choi, "Beam selection in mm-wave multiuser MIMO systems using compressive sensing," *IEEE Trans. Commun.*, vol. 63, no. 8, pp. 2936–2947, Aug. 2015.

[21] J. Nam, A. Adhikary, J.-Y. Ahn, and G. Caire, "Joint spatial division and multiplexing: Opportunistic beamforming, user grouping and simplified downlink scheduling," *IEEE J. Sel. Topics Signal Process.*, vol. 8, no. 5, pp. 876–890, Oct. 2014.

[22] S. Saur and H. Halbauer, "Exploring the vertical dimension of dynamic beam steering," in *Proc. 8th Int. Workshop Multi-Carrier Syst. Solutions (MC-SS)*, May 2011, pp. 1–5.

[23] N. Seifi, M. Coldrey, and T. Svensson, "Throughput optimization in MU-MIMO systems via exploiting BS antenna tilt," in *Proc. IEEE Global Commun. Conf. (GLOBECOM) Workshops*, 2012, pp. 653–657.

[24] *Guidelines for Evaluation of Radio Interface Technologies for IMT-Advanced*, document ITU-R M.2135-1, Radiocommunication Sector of International Telecommunication Union, 2009.

[25] *Further Advancements for E-UTRA Physical Layer Aspects (Release 9)*, document 3GPP TR 36.814 V9.0.0, Mar. 2010.

[26] N. Seifi, J. Zhang, R. W. Heath, Jr., T. Svensson, and M. Coldrey, "Coordinated 3D beamforming for interference management in cellular networks," *IEEE Trans. Wireless Commun.*, vol. 13, no. 10, pp. 5396–5410, Oct. 2014.

[27] W. Lee, S.-R. Lee, H.-B. Kong, S. Lee, and I. Lee, "Downlink vertical beamforming designs for active antenna systems," *IEEE Trans. Commun.*, vol. 62, no. 6, pp. 1897–1907, Jun. 2014.

[28] A. M. Tulino, A. Lozano, and S. Verdú, "Impact of antenna correlation on the capacity of multiantenna channels," *IEEE Trans. Inf. Theory*, vol. 51, no. 7, pp. 2491–2509, Jul. 2005.

[29] S. K. Yong and J. S. Thompson, "Three-dimensional spatial fading correlation models for compact MIMO receivers," *IEEE Trans. Wireless Commun.*, vol. 4, no. 6, pp. 2856–2869, Nov. 2005.

[30] M. Shafi *et al.*, "Polarized MIMO channels in 3-D: Models, measurements and mutual information," *IEEE J. Sel. Areas Commun.*, vol. 24, no. 3, pp. 514–527, Mar. 2006.

[31] K. Mammasis, R. W. Stewart, and J. S. Thompson, "Spatial fading correlation model using mixtures of Von Mises Fisher distributions," *IEEE Trans. Wireless Commun.*, vol. 8, no. 4, pp. 2046–2055, Apr. 2009.

[32] P. D. Teal, T. D. Abhayapala, and R. A. Kennedy, "Spatial correlation for general distributions of scatterers," *IEEE Signal Process. Lett.*, vol. 9, no. 10, pp. 305–308, Oct. 2002.

[33] Q.-U.-A. Nadeem, A. Kammoun, M. Debbah, and M.-S. Alouini, "A generalized spatial correlation model for 3D MIMO channels based on the Fourier coefficients of power spectrums," *IEEE Trans. Signal Process.*, vol. 63, no. 14, pp. 3671–3686, Jul. 2015.

[34] D. J. Hofsommer and M. L. Potters, "Table of Fourier coefficients of associated Legendre functions," Comput. Dep. Math. Center, Amsterdam, The Netherlands, Tech. Rep. 478, Jun. 1960.

[35] Q.-U.-A. Nadeem, A. Kammoun, M. Debbah, and M.-S. Alouini, "Spatial correlation characterization of a full dimension massive MIMO system," in *Proc. IEEE Global Commun. Conf. (GLOBECOM)*, Dec. 2016, pp. 1–7.

[36] M. R. McKay and I. B. Collings, "On the capacity of frequency-flat and frequency-selective Rician MIMO channels with single-ended correlation," *IEEE Trans. Wireless Commun.*, vol. 5, no. 8, pp. 2038–2043, Aug. 2006.

[37] B. Partov, D. J. Leith, and R. Razavi, "Utility fair optimization of antenna tilt angles in LTE networks," *IEEE/ACM Trans. Netw.*, vol. 23, no. 1, pp. 175–185, Feb. 2015.

[38] R. Couillet and M. Debbah, *Random Matrix Methods for Wireless Communications*, 1st ed. New York, NY, USA: Cambridge Univ. Press, 2011.

[39] T. K. Y. Lo, "Maximum ratio transmission," *IEEE Trans. Commun.*, vol. 47, no. 10, pp. 1458–1461, Oct. 1999.

[40] E. Karipidis, N. D. Sidiropoulos, and Z.-Q. Luo, "Quality of service and max-min fair transmit beamforming to multiple cochannel multicast groups," *IEEE Trans. Signal Process.*, vol. 56, no. 3, pp. 1268–1279, Mar. 2008.

[41] A. Zappone and E. Jorswieck, "Energy efficiency in wireless networks via fractional programming theory," *Found. Trends Commun. Inf. Theory*, vol. 11, nos. 3–4, pp. 185–396, 2015.

[42] A. Zappone, L. Sanguinetti, G. Bacci, E. Jorswieck, and M. Debbah, "Energy-efficient power control: A look at 5G wireless technologies," *IEEE Trans. Signal Process.*, vol. 64, no. 7, pp. 1668–1683, Apr. 2016.

[43] K. Schäcke, "On the Kronecker product," M.S. thesis, Dept. Math., Univ. Waterloo, Waterloo, ON, Canada, 2004.

[44] Z.-Q. Luo, W.-K. Ma, A. M.-C. So, Y. Ye, and S. Zhang, "Semidefinite relaxation of quadratic optimization problems," *IEEE Signal Process. Mag.*, vol. 27, no. 3, pp. 20–34, May 2010.

[45] J. Shao, *Mathematical Statistics*. New York, NY, USA: Springer, 2007.

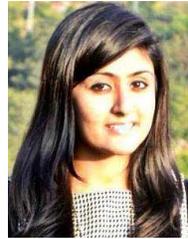

**Qurrat-Ul-Ain Nadeem** (S'15) was born in Lahore, Pakistan. She received the B.S. degree in electrical engineering from the Lahore University of Management Sciences, Pakistan, in 2013, and the M.S. degree in electrical engineering from the King Abdullah University of Science and Technology, Thuwal, Saudi Arabia, in 2015, where she is currently pursuing the Ph.D. degree with the Electrical Engineering Department. Her research interests include channel modeling, antenna array design and performance analysis of wireless communication systems.

**Abla Kammoun** (M'10) was born in Sfax, Tunisia. She received the Engineering degree in signal and systems from the Tunisia Polytechnic School, La Marsa, and the master's and Ph.D. degrees in digital communications from Telecom Paris Tech. From 2010 to 2012, she was a Post-Doctoral Researcher with the TSI Department, Telecom Paris Tech. She was then with the Alcatel-Lucent Chair on Flexible Radio, Suplec, until 2013. She is currently a Research Scientist with the King Abdullah University of Science and Technology. Her research interests include performance analysis, random matrix theory, and semi-blind channel estimation.

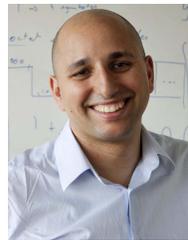

**Mérouane Debbah** (S'01–AM'03–M'04–SM'08–F'15) received the M.Sc. and Ph.D. degrees from the Ecole Normale Suprieure de Cachan, France. He was with Motorola Labs, Saclay, France, from 1999 to 2002, and the Vienna Research Center for Telecommunications, Vienna, Austria, until 2003. From 2003 to 2007, he was with the Mobile Communications Department, Institut Eurecom, Sophia Antipolis, France, as an Assistant Professor. Since 2007, he has been a Full Professor with CentraleSupelec, Gif-sur-Yvette, France. From 2007 to 2014, he was the Director of the Alcatel-Lucent Chair on Flexible Radio. Since 2014, he has been the Vice-President of the Huawei France R&D Center and the Director of the Mathematical and Algorithmic Sciences Laboratory. His research interests lie in fundamental mathematics, algorithms, statistics, and information and communication sciences research. He is a WWRF Fellow and a member of the academic senate of Paris–Saclay. He was a recipient of the ERC grant MORE (Advanced Mathematical Tools for Complex Network Engineering). He received 17 best paper awards, among which the 2007 IEEE GLOBECOM Best Paper Award, the Wi-Opt 2009 Best Paper Award, the 2010 Newcom++ Best Paper Award, the WUN CogCom Best Paper 2012 and 2013 Award, the 2014 WCNC Best Paper Award, the 2015 ICC Best Paper Award, the 2015 IEEE Communications Society Leonard G. Abraham Prize, the 2015 IEEE Communications Society Fred W. Ellersick Prize, the 2016 IEEE Communications Society Best Tutorial Paper Award, the 2016 European Wireless Best Paper Award, the 2017 EURASIP Best Paper Award, and the Valuetools 2007, Valuetools 2008, CrownCom2009, Valuetools 2012, and SAM 2014 best student paper awards. He was a recipient of the Mario Boella Award in 2005, the IEEE Glavieux Prize Award in 2011, and the Qualcomm Innovation Prize Award in 2012. He is an Associate Editor in Chief of the journal *Random Matrix: Theory and Applications.* He was an Associate and Senior Area Editor of the IEEE Transactions on Signal Processing from 2011 to 2013 and 2013 to 2014, respectively.

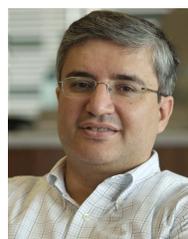

**Mohamed-Slim Alouini** (S'94–M'98–SM'03–F'09) was born in Tunis, Tunisia. He received the Ph.D. degree in electrical engineering from the California Institute of Technology, Pasadena, CA, USA, in 1998. He served as a Faculty Member with the University of Minnesota, Minneapolis, MN, USA, then Texas A&M University at Qatar, Doha, Qatar, before joining King Abdullah University of Science and Technology, Thuwal, Saudi Arabia, as a Professor of electrical engineering in 2009. His current research interests include the modeling, design, and performance analysis of wireless communication systems.